\documentclass[aps,onecolumn,showpacs]{revtex4}
\usepackage{graphicx}
\usepackage{epsfig}
\usepackage{amssymb}
\usepackage{graphicx}
\usepackage{amsfonts}
\usepackage{latexsym}
\usepackage{amsmath}
\usepackage{subfigure}
\usepackage{mathrsfs}
\numberwithin{equation}{section}

\newcommand\encadremath[1]{\vbox{\hrule\hbox{\vrule\kern8pt
\vbox{\kern8pt \hbox{$\displaystyle #1$}\kern8pt}
\kern8pt\vrule}\hrule}}
\def\enca#1{\vbox{\hrule\hbox{
\vrule\kern8pt\vbox{\kern8pt \hbox{$\displaystyle #1$} \kern8pt}
\kern8pt\vrule}\hrule}}

\newcommand{\tpsi}
{\widehat\Psi}
\newcommand{\tphi}
{\widehat\Phi}
\newcommand{\mnb}
{\prod_M^{\overline N}}
\newcommand{\nbnn}
{\prod_{N+1}^{\overline N}}
\newcommand{\mn}
{\prod_M^N}

\newcommand{\mdown}
{\prod_M}
\newcommand{\nup}
{\prod^N}
\newcommand{\ndown}
{\prod_N}
\newcommand{\nbup}
{\prod^{\overline N}}
\newcommand{\nbdown}
{\prod_{\overline N}}
\newcommand{\nnb}
{\prod^{\overline N}_N}

\newcommand{\bt}{(\beta)}

\def\pre#1#2#3{Phys. Rev. E  {#1} #2 #3}
\def\jpa#1#2#3{J. Phys. A {\bf #1}, #2 (#3)}

\def\cmp#1#2#3{Commun. Math. Phys. {\bf #1}, #2 (#3)}
\begin{document}
\title{Skew-orthogonal polynomials, differential systems and random matrix theory}
\author{Saugata Ghosh}\email{sghosh@ictp.it}
\affiliation{The Abdus Salam ICTP, Strada Costiera 11, 34100,
Trieste, Italy.}
\date{\today}
\begin{abstract}
We study skew-orthogonal polynomials with respect to the weight
function $\exp[-2V(x)]$, with
$V(x)=\sum_{K=1}^{2d}(u_{K}/{K})x^{K}$, $u_{2d}
> 0$, $d > 0$. A finite subsequence of such
skew-orthogonal polynomials arising in the study of Orthogonal and
Symplectic ensembles of random matrices, satisfy a system of
differential-difference-deformation equation. The vectors formed
by such subsequence has the rank equal to the degree of the
potential in the quaternion sense.  These solutions satisfy
certain  compatibility condition and hence admit a simultaneous
fundamental system of solutions.
\end{abstract}
\maketitle

\section{Introduction}
\subsection{Random matrices}
The concept of `Universality' in random matrix theory and its
various applications in real physical systems have attracted both
mathematicians and physicists in the last few decades
\cite{guhr,beenakker,ghosh,ghoshpandey,ghosh3,deift1,deift2,deeift3,baik,bleher,deift4,deift5,deift6,deift7}.
From mathematical point of view, the study of `Universality' in
the energy level correlations of random matrices require a good
understanding of the asymptotic behavior of certain families of
polynomials. For example Unitary ensembles of random matrices
require an understanding of the large $n$ behavior of orthogonal
(the one-matrix model) and bi-orthogonal (the two matrix model)
polynomials, while Orthogonal and Symplectic ensembles require
that of the skew-orthogonal polynomials. The rich literature
available on orthogonal polynomials
\cite{szego,deift1,deift2,deeift3,baik,bleher,deift4,deift5,deift6,deift7,eynard1,eynard11,plancherel}
and bi-orthogonal polynomials
\cite{eynard2,eynard3,eynard4,Kap,KM} have contributed a lot in
our understanding of the Unitary ensembles. Our aim is to develop
the theory of skew-orthogonal polynomials
\cite{ghosh,ghoshpandey,ghosh3,dyson1,mehta,mehta1,eynard} so that
we can have further insight into the one-matrix model for
Orthogonal and Symplectic ensembles of random matrices.

In this direction, previous experience with orthogonal and
bi-orthogonal polynomials arising in the two-matrix model
 makes us believe
that perhaps the most logical and rigorous way to study the
asymptotic properties of these skew-orthogonal polynomials is to
do the following: (1). First, we must look for a finite
subsequence of skew-orthonormal vectors which will satisfy a
mutually compatible system of differential-difference-deformation
equation. This requires one to fix the `size' or rank of this
finite sub-sequence. The Generalized Christoffel-Darboux formula
helps in this regard as it gives an indication of the number of
terms near the so-called `Fermi-level' which are actually required
to study this system. (2). The next step is to look for a mutually
compatible system of differential-difference-deformation equation
satisfied by the finite subsequence of these vectors and hence
find the so-called `fundamental solution'. (3). Finally, to
formulate a quaternion matrix Riemann-Hilbert problem and by
applying the steepest descent method, obtain the
Plancherel-Rotach-type \cite{plancherel} formula for
skew-orthogonal polynomials.

Reference \cite{ghosh3} achieved the first goal. In this paper, we
try to address the second issue of the existence of the
`fundamental solution' of a system of
differential-difference-deformation equation. One of the
by-products is that we observe certain duality property between
the skew-orthonormal vectors of Orthogonal and Symplectic
ensembles. (In fact, this justifies further the use of
skew-orthogonal polynomials to ordinary orthogonal polynomials in
studying these ensembles
\cite{stoj1,stoj1a,stoj2,deift1,deift2,nagao,tracy1,tracy2,tracy3,adler,widom}.)
The third part, which is also the most crucial one, is not yet
fully understood.

We shall consider an ensemble of $2N$ dimensional  matrices $H$
with probability distribution
\begin{equation}
P_{\beta,N}(H)dH=\frac{1}{{\cal Z}_{\beta N}}\exp[-[2{\rm
Tr}V(H)]]dH,
\end{equation}
where the parameter $\beta=1$ and $4$ corresponds to the joint
probability density for Orthogonal and Symplectic ensembles of
random matrices. The ``potential'' $V(x)$ is a polynomial of
degree $2d$ with positive leading coefficient:
\begin{equation}
\label{V(x)} V(x)=\sum_{K=1}^{2d}\frac{u_{K}}{K}x^{K},
\hspace{0.5cm} u_{2d} > 0, \hspace{0.5cm} d > 0,
\end{equation}
where $u_{K}$ is called the deformation parameter. ${\cal
Z}_{\beta N}$ is the so called `partition function':

\begin{equation}
{\cal Z}_{\beta N}:= \int_{H\in M^{\bt}_{2N}}\exp[-[2{\rm
Tr}V(H)]]dH=N!\prod_{j=0}^{N-1}g_{j}^{\bt},
\end{equation}
where $M^{\bt}_{2N}$ is a set of all $2N\times 2N$ real symmetric
($\beta=1$) and quaternion real self dual ($\beta=4$) matrices.
$dH$ is the standard Haar measure. $g_j^{\bt}$ is the
skew-normalization constant for  polynomials related to Orthogonal
and Symplectic ensembles \cite{dyson1}.

{\it Remark on notation.} Before entering the details of
calculation, we must mention that throughout this paper, we have
followed to a great extent notations used in
\cite{eynard,eynard1,eynard2,eynard3,eynard4}. Apart from having
the advantage of using a well established and compact notation,
this strategy will also highlight the striking similarity (as well
as the difference) between skew-orthogonal polynomials and their
bi-orthogonal counterpart.



\subsection{Skew-Orthogonal Polynomials: relevance in Orthogonal and Symplectic
ensembles}


{\bf Definition}:

For Orthogonal and Symplectic ensembles of random matrices, we
define semi-infinite vectors:
\begin{eqnarray}
\label{phiinfty}
\Phi^{\bt}(x)={({\Phi_0^{\bt}}^t(x)\ldots{\Phi_n^{\bt}}^t(x)\ldots)}^t,\hspace{1cm}
\widehat{\Phi}^{\bt}(x)=
-{\Phi^{\bt}}^{t}(x)Z,\\
\label{psiinfty}
\Psi^{\bt}(x)={({\Psi_0^{\bt}}^t(x)\ldots{\Psi_n^{\bt}}^t(x)\ldots)}^t,\hspace{1cm}
\widehat{\Psi}^{\bt}(x)= -{\Psi^{\bt}}^{t}(x)Z,
\end{eqnarray}
where
\begin{equation}
\label{psi} \Psi^{(4)}_{n}(x) =
  \Phi '^{(4)}_{n}(x),\hspace{0.3cm}
\Psi^{(1)}_{n}(x)=\int_{\mathbb
R}\Phi^{(1)}_{n}(y)\epsilon(x-y)dy.
\end{equation}
Each entry in these semi-infinite vectors is a $(2\times 1)$
matrix:
\begin{eqnarray}
\Phi^{(\beta)}_n(x) = \left(\begin{array}{c}
\phi^{(\beta)}_{2n}(x)      \\
\phi^{(\beta)}_{2n+1}(x)   \\
\end{array}\right),
\hspace{1cm} \Psi^{(\beta)}_n(x) = \left(\begin{array}{c}
\psi^{(\beta)}_{2n}(x)      \\
\psi^{(\beta)}_{2n+1}(x)   \\
\end{array}\right),
\end{eqnarray}
where
\begin{eqnarray}
\label{quasipolynomial} \phi_{n}^{\bt}(x) &=&
\frac{1}{\sqrt{g^{\bt}_n}}\pi^{\bt}_{n}(x)\exp[-V(x)],
\end{eqnarray}
is the weighted skew-normalized polynomial (often called the
quasi-polynomial) and
\begin{eqnarray}
\label{monopoly}
 \pi^{\bt}_{n}(x) &=&
\sum^{n}_{k=0}c^{(n,\beta)}_{k}x^k,{\hspace{1cm}c^{(n,\beta)}_{n}=1},{\hspace{1cm}}n\in\mathbb{N},
\end{eqnarray}
is monic skew-orthogonal polynomial of order $n$.
\begin{eqnarray}
Z &=& \left(\begin{array}{cc}
0 & 1     \\
-1 & 0         \\
\end{array}\right)
\dotplus \ldots \dotplus
\end{eqnarray}
is semi-infinite anti-symmetric block-diagonal matrix with
$Z^2=-1$ and
\begin{eqnarray}
\epsilon(r) &=& \frac{|r|}{r}
\end{eqnarray}
is the step function.
 These vectors satisfy skew-orthonormality
relation:
\begin{equation}
\label{ortho1} ({\Phi}^{\bt}_n,\tpsi_m^{\bt}) \equiv\int_{\mathbb
R}{\Phi}^{\bt}_n(x){\tpsi}_m^{\bt}(x) dx
=\delta_{nm}\left(\begin{array}{cc}
1 & 0     \\
0 & 1         \\
\end{array}\right),
\hspace{0.5cm} n,m\in\mathbb N.
\end{equation}
{\it Remark}: This definition differs from that of
\cite{mehta,ghosh,ghoshpandey} for $\beta=4$ by a factor of $2$,
which is incorporated in the normalization constant.

These skew-orthonormal vectors  form the necessary constituents of
the kernel functions required to study different statistical
properties of Orthogonal ensembles and Symplectic ensembles of
random matrices. For example, the two point correlation function
can be written in terms of the $2\times 2$ kernel function
\cite{mehta,mehta1}:

\begin{eqnarray}
{\sigma}^{(\beta)}_2(x,y) &=& \left(\begin{array}{cc}
S^{(\beta)}_{2N}(x,y) & D^{(\beta)}_{2N}(x,y)     \\
I^{(\beta)}_{2N}(x,y)+\frac{\delta_{1,\beta}}{2}\epsilon (x-y) & S^{(\beta)}_{2N}(y,x)         \\
\end{array}\right),
\end{eqnarray}
where $\delta$ is the kronecker delta and

\begin{equation}
\label{s} S^{(\beta)}_{2N}(x,y)=
{{\widehat{\Psi}}^{(\beta)}}(x)\prod_{N} \Phi^{(\beta)}(y),
\hspace{1cm} D^{(\beta)}_{2N}(x,y)=
{{\widehat{\Phi}}^{(\beta)}}(x)\prod_{N} \Phi^{(\beta)}(y),
\end{equation}

\begin{equation}
\label{i} I^{(\beta)}_{2N}(x,y)=
{{\widehat{\Psi}}^{(\beta)}}(x)\prod_{N} \Psi^{(\beta)}(y),
\hspace{1cm} S^{(\beta)}_{2N}(y,x)=
{{\widehat{\Phi}}^{(\beta)}}(x)\prod_{N} \Psi^{(\beta)}(y).
\end{equation}
The level density is given by
\begin{equation}
{\rho}^{(\beta)}(x,x)=S_{2N}^{(\beta)}(x,x).
\end{equation}
The matrix
\begin{equation}
\prod_{N}={\rm diag}(\underbrace{{\mathbf 1},\ldots,{\mathbf
1}}_{N},0,\ldots, 0)
\end{equation}
is formed by $N$ ($2\times 2$) unit matrices (i.e. a unit matrix
of size $2N$ in real space). In general, this will be used to
truncate semi-infinite matrices and in the above case, the
semi-infinite vectors of the finite sum in Eqs.(\ref{s}) and
(\ref{i}).

\subsection{The Generalised Christoffel Darboux sum}

Here, we present a summary of the main results of
Ref.\cite{ghosh3}, where we studied the kernel function
$S^{(\beta)}_{2N}(x,y)$. We expand $(x\Phi^{\bt}(x))$,
${(\Phi^{\bt}(x))}'$ and ${(x\Phi^{\bt}(x))}'$ in terms of
$\Phi^{(\beta)}(x)$ (and hence introduce the semi-infinite
matrices $Q^{\bt}$, $P^{\bt}$ and $R^{\bt}$ respectively):

\begin{eqnarray}
\label{Q} x\Phi^{\bt}(x) &=& Q^{\bt}\Phi^{\bt}(x),\\
 \label{PR4} \Psi^{(4)}(x) &=&
P^{(4)}\Phi^{(4)}(x),{\hspace{1cm}}
x\Psi^{(4)}(x)=R^{(4)}\Phi^{(4)}(x),\\
\label{PR1} 2\Phi^{(1)}(x) &=&
P^{(1)}\Psi^{(1)}(x),{\hspace{0.7cm}}2x\Phi^{(1)}(x)=R^{(1)}\Psi^{(1)}(x),
\end{eqnarray}
where Eq.(\ref{PR1}) is obtained by multiplying the above
expansion by $\epsilon(y-x)$ and integrating by parts. Here
\begin{eqnarray}
R^{(4)}=P^{(4)}Q^{(4)},\qquad R^{(1)}=Q^{(1)}P^{(1)}.
\end{eqnarray}

The matrices $P^{\bt}$, $Q^{\bt}$ and $R^{\bt}$ are quaternion
matrices. For a nice introduction to the subject, the reader is
referred to the book by Prof. M. L. Mehta \cite{mehta1}.

Defining the quaternion-matrix

\begin{equation}
{\overline R}^{(4)}(x):=R^{(4)}-xP^{(4)},\hspace{1cm}{\overline
R}^{(1)}(x):=\frac{1}{2}\left[R^{(1)}-xP^{(1)}\right]
\end{equation}
the Generalized Christoffel Darboux formula for the Symplectic
ensembles ($\beta=4$) can be written as:

\begin{eqnarray}
\label{GCD4}(x-y)S^{(4)}_{2N}(x,y)=-
{{\widehat{\Phi}}}^{(4)}(x)\left[{\overline
R}^{(4)}(x),\prod_{N}\right]{\Phi}^{(4)}(y).
\end{eqnarray}

For Orthogonal ensembles ($\beta=1$) the Generalized Christoffel
Darboux is given by:
\begin{eqnarray}
\label{GCD1} (y-x)S^{(1)}_{2N}(x,y)=
-{{\widehat{\Psi}}}^{(1)}(x)\left[{\overline
R}^{(1)}(y),\prod_{N}\right]\Psi^{(1)}(y).
\end{eqnarray}
The Generalized Christoffel-Darboux matrix $\left[{\overline
R}^{\bt}(x),\prod_{N}\right]$ in terms of the elements of the
quaternion matrix takes the form

\begin{eqnarray}
\label{rgcd} {[{\overline R}^{\bt}(x),\prod_{N}]}=
\left(\begin{array}{cccccc}
0&0&0&{\overline R}_{2N-2d,2N}^{\bt}(x)&0&0 \\
0&0&0&\vdots&\ddots&0 \\
0&0&0&{\overline R}_{2N-1,2N}^{\bt}(x)&\ldots
                     &{\overline R}_{2N-1,2N+2d-1}^{\bt}(x) \\
-{\overline R}_{2N,2N-2d}^{\bt}(x) &\ldots
&-{\overline R}_{2N,2N-1}^{\bt}(x)&0&0&0\\
0&\ddots&\vdots&\vdots & 0&0 \\
0&0&-{\overline R}_{2N+2d-1,2N-1}^{\bt}(x)&0& \ldots & 0 \\
\end{array}\right),N\geq d.
\end{eqnarray}

Here, we must point out that there is a small difference with the
notation used in the second part of Ref.\cite{ghosh3}, where we
used a matrix of size $2N+2$ to prove the `Universality' in the
Gaussian case. Also the definition of $R^{(1)}$ and $P^{(1)}$
differs by a factor of $2$.


\subsection{Difference-differential-deformation equations}


A quick glance at Eq.(\ref{rgcd}) reveals that the relevant
vectors contributing to the correlation function are
$\phi^{\bt}_{(2N-2d)}(x)\ldots \phi^{\bt}_{(2N+2d-1)}(x)$. This
prompts us to define a finite subsequence (or window:
$W\Rrightarrow \{N-d,\ldots,N+d-1\}, N\geq d$) of skew-orthogonal
vectors of size $2d$ (in the quaternionic sense),
${{\Phi}^{\bt}_W}(x)$ and ${{{\Psi}}_W^{\bt}}(x)$:
\begin{eqnarray}
\label{window} \Phi^{(\beta)}_{W}(x):= \left(\begin{array}{c}
\phi^{\bt}_{2N-2d}{(x)}      \\
\vdots     \\
\phi^{\bt}_{2N+2d-1}{(x)}               \\
\end{array}\right)\equiv
\left(\begin{array}{c}
\Phi^{\bt}_{{N-d}}{(x)}      \\
\vdots     \\
\Phi^{\bt}_{N+d-1}{(x)}               \\
\end{array}\right),\hspace{1cm}N\geq d,\\
\label{window1} \Psi^{(\beta)}_{W}(x):= \left(\begin{array}{c}
\psi^{\bt}_{2N-2d}{(x)}      \\
\vdots     \\
\psi^{\bt}_{2N+2d-1}{(x)}               \\
\end{array}\right)\equiv
\left(\begin{array}{c}
\Psi^{\bt}_{{N-d}}{(x)}      \\
\vdots     \\
\Psi^{\bt}_{N+d-1}{(x)}               \\
\end{array}\right),\hspace{1cm}N\geq d,
\end{eqnarray}
such that
\begin{eqnarray}
\Phi_{W\pm j}^{\bt}(x):={\left({\Phi_{N-d\pm j}^{{\bt}^t}}(x)
\ldots \Phi_{N+d-1 \pm
j}^{{\bt}^t}(x)\right)}^{t},{\hspace{1cm}}\forall j\in \mathbb N,\\
\Psi_{W\pm j}^{\bt}(x):={\left(\Psi_{N-d\pm j}^{{\bt}^t}(x) \ldots
\Psi_{N+d-1 \pm j}^{{\bt}^t}(x)\right)}^{t},{\hspace{1cm}}\forall
j\in \mathbb N.
\end{eqnarray}
The rank of the window is equal to the degree of the potential
$V(x)$ in the quaternionic space and twice that of the degree of
$V(x)$ in the real space.

The recursion relations connecting the finite subsequence (or
window) with the upper or lower one is through the ladder operator
\begin{eqnarray}
{{\Phi}_{W+1}^{(4)}}(x)=A_N^{(4)}(x){{\Phi}_W^{(4)}}(x),
\hspace{1cm}
{{\Phi}_{W-1}^{(4)}}(x)={(A_{N-1}^{(4)}}(x))^{-1}{{\Phi}_W^{(4)}}(x),
\end{eqnarray}
\begin{eqnarray}
{\Psi}_{W+1}^{(1)}(x)=A_N^{(1)}(x){{\Psi}_W^{(1)}}(x),
\hspace{1cm}
{\Psi}_{W-1}^{(1)}(x)={(A_{N-1}^{(1)}}(x))^{-1}{\Psi}_W^{(1)}(x),
\end{eqnarray}
and
\begin{eqnarray}
{{\Psi}_{W+1}^{(4)}}(x)=B_N^{(4)}(x){{\Psi}_W^{(4)}}(x),
\hspace{1cm}
{{\Psi}_{W-1}^{(4)}}(x)={(B_{N-1}^{(4)}}(x))^{-1}{{\Psi}_W^{(4)}}(x),
\end{eqnarray}
\begin{eqnarray}
{{\Phi}_{W+1}^{(1)}}(x)=B_{N}^{(1)}(x){{\Phi}_W^{(1)}}(x),
\hspace{1cm}
{{\Phi}_{W-1}^{(1)}}(x)={(B_{N-1}^{(1)}}(x))^{-1}{{\Phi}_W^{(1)}}(x),
\end{eqnarray}

The vectors ${{\Phi}_W^{\bt}}(x)$ and ${{\Psi}_W^{\bt}}(x)$ also
satisfy a system of ODEs:

\begin{eqnarray}
\nonumber \frac{d}{dx}\Psi^{(4)}_{W}(x)
                             = {\underline D}^{(4)}_{N}(x)\Psi^{(4)}_{W}(x), &&
\hspace{0.5cm} \vline \hspace{0.5cm} \frac{d}{dx}\Phi^{(1)}_{W}(x)
=
{\underline D}^{(1)}_{N}(x)\Phi^{(1)}_{W}(x),\\
\hline
\frac{d}{dx}\Phi^{(4)}_{W}(x) = D^{(4)}_{N}(x)\Phi^{(4)}_{W}(x),
&& \hspace{0.5cm} \vline \hspace{0.5cm}
\frac{d}{dx}\Psi^{(1)}_{W}(x) = {D}^{(1)}_{N}(x)\Psi^{(1)}_{W}(x).
\end{eqnarray}

Under an infinitesimal change of the deformation parameter $u_K$
(the coefficients of the polynomial potential $V(x)$), these
vectors satisfy a system of PDE's given by
\begin{eqnarray}
\nonumber \frac{\partial}{\partial u_K}\Psi^{(4)}_{W}(x)
                             = {{\underline U}_K^N}^{(4)}(x)\Psi^{(4)}_{W}(x), &&
\hspace{0.5cm} \vline \hspace{0.5cm} \frac{\partial}{\partial
u_K}\Phi^{(1)}_{W}(x) =
{\underline U_K^N}^{(1)}(x)\Phi^{(1)}_{W}(x),\\
\hline
\frac{\partial}{\partial u_K}\Phi^{(4)}_{W}(x) =
{U_K^N}^{(4)}(x)\Phi^{(4)}_{W}(x), && \hspace{0.5cm} \vline
\hspace{0.5cm} \frac{\partial}{\partial u_K}\Psi^{(1)}_{W}(x) =
{{U}_K^N}^{(1)}(x)\Psi^{(1)}_{W}(x).
\end{eqnarray}
Thus the finite subsequence of skew-orthogonal vectors satisfy a
system of difference-differential-deformation equation.

{\it Remark.} The two pairs of matrices ${\underline
D}^{(4)}_{N}(x)$ and ${\underline D}^{(1)}_{N}(x)$ (similarly
($D^{(1)}_{N}(x)$ and ${D}^{(4)}_{N}(x))$, ${\underline
U_K^N}^{(4)}(x)$ and ${\underline U_K^N}^{(1)}(x)$ (similarly
(${U_K^N}^{(1)}(x)$ and ${U_K^N}^{(4)}(x))$ and $A_N^{(4)}(x)$ and
$A_N^{(1)}(x)$ (similarly $B_N^{(4)}(x)$ and $B_N^{(1)}(x)$) are
dual in the sense that they remain invariant under an interchange
of $\Psi^{(4)}(x)\mapsto \Phi^{(1)}(x)$ and $Q^{(4)}\mapsto
Q^{(1)}$ (and similarly for $\Psi^{(1)}(x)$ and $\Phi^{(4)}(x)$).


\subsection{Compatibility}
The existence of recursion relations, differential equations and
deformation equations for vectors arising in the Orthogonal
ensembles and Symplectic ensembles of random matrices can be
viewed as just a projection of the semi-infinite functions
$\Phi^{\bt}(x)$ and $\Psi^{\bt}(x)$ onto the finite ``window''
$\Phi^{\bt}_{W}(x)$ and $\Psi^{\bt}_{W}(x)$ respectively. However,
we may also consider these equations as defining an overdetermined
system of finite difference-differential-deformation equations of
the vector functions and see that these systems are compatible.
This leads to the existence of a  fundamental matrix solution,
denoted by $\Phi_{[W]}^{\bt}(x)$ and $\Psi_{[W]}^{\bt}(x)$ where
all the column vectors satisfy the above
difference-differential-deformation equations simultaneously.

The compatibility of the deformation and difference equation with
the differential equation imply that the generalized monodromy of
the operator $\left(\frac{d}{dx}-D_{N}^{(\beta)}(x)\right)$ and
$\left(\frac{d}{dx}-{\underline D}_{N}^{(\beta)}(x)\right)$ is
invariant under $u_K$ deformations and shifts in $N$.



\vspace{1cm} {\bf Outline of the article}



-Section 2 deals with the properties of different finite band
matrices related to skew-orthogonal polynomials.

-In section 3, we study the system of PDE's arising from the
infinitesimal change of the deformation parameter $u_K$.

-We derive the difference relations satisfied by the finite
subsequence of skew-orthogonal vectors in section 4.

-In section 5, we derive the folding function which is used to
project any given vector $\Phi^{\bt}_{n}$ (or $\Psi^{\bt}_{n}$)
onto its finite subsequence or window.

-In section 6,  we obtain the folded deformation matrix,
 using the results of section 5.

- The differential equation for skew-orthonormal vectors, using
the results obtained in Section 6, is derived in section 7.

-In section 8, we discuss the existence of the Cauchy-like
transforms of the skew-orthogonal vectors of order $n$ as the
other solutions to the differential-deformation-difference
equations, for fairly large $n$.

 -We prove compatibility conditions for these difference-deformation-differential system
 in section 9.



\section{Recursion Relations and finite band matrices.}

In this section, we will study in detail the different recursion
relations satisfied by the skew-orthonormal vectors. Here, we must
note that unlike the orthogonal polynomials, skew-orthonormal
vectors do not satisfy a three-term recursion relation and hence
do not give birth to the tri-diagonal Jacobi matrix.

Differentiating Eqs.(\ref{PR4})-(\ref{PR1}), we get

\begin{eqnarray}
\label{recursion} \frac{d}{dx}{\Phi^{\bt}(x)} =
P^{\bt}{\Phi^{\bt}(x)},\qquad
\frac{d}{dx}{\Psi^{\bt}(x)} = P^{\bt}{\Psi^{\bt}(x)},\qquad \beta=1,4,\\
\label{recursion4} x\frac{d}{dx}{\Phi^{(4)}(x)} =
R^{(4)}{\Phi^{(4)}(x)},\qquad
x\frac{d}{dx}{\Psi^{(4)}(x)} = \left[R^{(4)}-{\bf 1}\right]{\Psi^{(4)}(x)},\\
\label{recursion5} x\frac{d}{dx}{\Psi^{(1)}(x)} =
R^{(1)}{\Psi^{(1)}(x)},\qquad x\frac{d}{dx}{\Phi^{(1)}(x)} =
\left[R^{(1)}-{\bf 1}\right]{\Phi^{(1)}(x)},
\end{eqnarray}
where $\Phi^{\bt}(x)$ and $\Psi^{\bt}(x)$ are semi-infinite column
vectors defined in Eqs.(\ref{phiinfty},\ref{psiinfty}). This is
equivalent to saying
\begin{eqnarray}
\label{beta4} [R^{(4)}-xP^{(4)}]\Phi^{(4)}(x) &=&
0,{\hspace{0.7cm}}
[R^{(4)}-xP^{(4)}-{\bf 1}]\Psi^{(4)}(x) = 0,\\
\label{beta1} [R^{(1)}-xP^{(1)}]\Psi^{(1)}(x) &=&
0,{\hspace{0.7cm}} [R^{(1)}-xP^{(1)}-{\bf 1}]\Phi^{(1)}(x) = 0.
\end{eqnarray}

These finite band matrices satisfy the following commutation
relations:

\begin{equation}
\label{string} [Q^{(\beta)},P^{(\beta)}]={\bf 1},\hspace{1cm}
[R^{(\beta)},P^{(\beta)}]=P^{(\beta)}.
\end{equation}

Here, each entry is a $2\times 2$ quaternion of the form:
\begin{eqnarray}
A_{n,m}:= \left[\begin{array}{cc} {\tilde A}_{2n,2m} &
{\tilde A}_{2n,2m+1}  \\
{\tilde A}_{2n+1,2m}     & {\tilde A}_{2n+1,2m+1}  \\
\end{array}\right].
\end{eqnarray}

Using  $(\psi_{n}^{(4)},\psi_{m}^{(4)})$ and
$(x\psi_{n}^{(4)},\psi_{m}^{(4)})$ for $\beta=4$, and replacing
$\psi^{(4)}(x)$ by $\phi^{(1)}(x)$ for $\beta=1$, we get
\begin{eqnarray}
\label{dual}
P^{\bt}=-{P^{\bt}}^D,\hspace{1cm}R^{\bt}=-{R^{\bt}}^D,
\end{eqnarray}
where dual of a matrix $A$ is defined as \cite{mehta1}:
\begin{equation}
A^D=-ZA^{t}Z.
\end{equation}

{\it Remark.} The condition of anti-self duality imposes a much
lighter restriction on a diagonal quaternion than antisymmetry
condition on a diagonal matrix element. For example, it leaves the
off-diagonal entry of the diagonal quaternion arbitrary. This is
the reason why the odd skew-orthogonal polynomials are arbitrary
upto the addition of a lower even polynomial.

However starting with $(x\phi_{n}^{\bt},\psi_{m}^{\bt})$ and using
(\ref{string}), we get
\begin{eqnarray}
(Q^{\bt}-{Q^{\bt}}^D)P^{\bt}=P^{\bt}(Q^{\bt}-{Q^{\bt}}^D)={\bf 1}.
\end{eqnarray}
As pointed out in the beginning of the section, this essentially
means that like the orthogonal polynomials, we do not have a
tri-diagonal Jacobi-matrix for skew-orthogonal polynomials. It is
this relation that  causes the Generalized Christoffel-Darboux sum
to have a local behavior (i.e. the rank has dependence on the
measure or weight function). In a sense, this is a major setback
in our hope of defining a $2\times 2$ matrix Riemann-Hilbert
problem for  skew-orthogonal vectors, similar to that of the
orthogonal polynomials. Also from the dependence of $P^{\bt}$ and
$R^{\bt}$ on $2d$ (the degree of the polynomial $V(x)$), we can
conclude that the `size' of our q-matrix Riemann-Hilbert problem
will depend on $d$.

From \cite{ghosh3}, we also have

\begin{equation}
\label{p+} P^{(\beta)}+[{V'(Q^{(\beta)})]}={\rm
lower},\hspace{1cm}
 R^{(\beta)}+[{Q^{(\beta)}V'(Q^{(\beta)})]}={\rm
lower}_+,
\end{equation}
where `lower' and `${\rm lower}_+$' denotes a strictly lower
triangular matrix and a lower triangular matrix with the principal
diagonal respectively.

Having obtained the recursion relations for the skew-orthonormal
vectors, we introduce a convenient notation to express
band-matrices $P^{\bt}$ and $R^{\bt}$:
\begin{equation}
\sum_{m=n-d}^{n+d}
P^{\bt}_{nm}\Phi^{\bt}_{m}(x)=\sum_{k=-d}^{d}\zeta^{\bt}_{k}(n)\Phi^{\bt}_{n-k}(x),\hspace{1cm}
\sum_{m=n-d}^{n+d}
R^{\bt}_{nm}\Phi^{\bt}_{m}(x)=\sum_{k=-d}^{d}\eta^{\bt}_{k}(n)\Phi^{\bt}_{n-k}(x),
\end{equation}
and
\begin{equation}
\label{alpha} \sum_{m=n-d}^{n+d}
R^{\bt}_{nm}\Phi^{\bt}_{m}(x)-x\sum_{m=n-d}^{n+d}P^{\bt}_{nm}\Phi^{\bt}_{m}(x)
=\sum_{k=-d}^{d}\alpha^{\bt}_{k}(n)\Phi^{\bt}_{n-k}(x),
\end{equation}

where
\begin{equation}
{\eta^{\bt}_{k}}:={\rm
diag}({\eta^{\bt}_{k}}(0),\ldots,{\eta^{\bt}_{k}}(n),\ldots),\qquad
{\zeta^{\bt}_{k}}:={\rm
diag}({\zeta^{\bt}_{k}}(0),\ldots,{\zeta^{\bt}_{k}}(n),\ldots),\qquad
{\alpha^{\bt}_{k}}:={\rm
diag}({\alpha^{\bt}_{k}}(0),\ldots,{\alpha^{\bt}_{k}}(n),\ldots).
\end{equation}
Here, we have suppressed the $x$-dependence of
${\alpha^{\bt}_{k}}(n)$. This notation will be useful in deriving
the folding function.

 More
explicitly, with this notation, we can write
\begin{eqnarray}
{P^{\bt}}= \left[\begin{array}{cccccccc} \zeta^{\bt}_{0}(0) &
\zeta^{\bt}_{-1}(0) & \ldots  & \ldots &
\zeta^{\bt}_{-d}(0)  & 0 & 0  & \ldots \\
\zeta^{\bt}_{1}(1)     & \zeta^{\bt}_{0}(1) & \zeta^{\bt}_{-1}(1)
& \ldots &
\zeta^{\bt}_{-d+1}(1)  & \zeta^{\bt}_{-d}(1) & 0  & \ldots \\
\zeta^{\bt}_{2}(2)     & \zeta^{\bt}_{1}(2) & \zeta^{\bt}_{0}(2) &
\ldots &
\zeta^{\bt}_{-d+2}(2)  & \zeta^{\bt}_{-d+1}(2) & \zeta^{\bt}_{-d}(2) & 0 \\
\vdots            & \vdots          & \vdots   & \ldots &
\ldots  & \ldots & \ldots  & \ddots \\
\zeta^{\bt}_{d}(d)     & \zeta^{\bt}_{d-1}(d) & \ldots & \ldots &
\zeta^{\bt}_{0}(d)  & \ldots & \ldots  & \ldots \\
0 & \zeta^{\bt}_{d}(d+1) & \ddots  & \ddots & \ddots  & \ddots & \ddots  & \ldots \\
0 & 0 & \zeta^{\bt}_{d}(d+2)  & \ddots  & \ddots & \ddots  &  \\
\end{array}\right),
\end{eqnarray}
and

\begin{eqnarray}
{R^{\bt}}= \left[\begin{array}{cccccccc} \eta^{\bt}_{0}(0) &
\eta^{\bt}_{-1}(0) & \ldots  & \ldots &
\eta^{\bt}_{-d}(0)  & 0 & 0  & \ldots \\
\eta^{\bt}_{1}(1)     & \eta^{\bt}_{0}(1) & \eta^{\bt}_{-1}(1)  &
\ldots &
\eta^{\bt}_{-d}(1)  & \eta^{\bt}_{-d}(1) & 0  & \ldots \\
\eta^{\bt}_{2}(2)     & \eta^{\bt}_{1}(2) & \eta^{\bt}_{0}(2)  &
\ldots &
\eta^{\bt}_{-d+1}(2)  & \eta^{\bt}_{-d}(2) & \eta^{\bt}_{-d}(2) & 0 \\
\vdots            & \vdots          & \vdots   & \ldots &
\ldots  & \ldots & \ldots  & \ddots \\
\eta^{\bt}_{d}(d)     & \eta^{\bt}_{d-1}(d) & \ldots & \ldots &
\eta^{\bt}_{0}(d)  & \ldots & \ldots  & \ldots \\
0 & \eta^{\bt}_{d}(d+1) & \ddots  & \ddots & \ddots  & \ddots & \ddots  & \ldots \\
0 & 0 & \eta^{\bt}_{d}(d+2)  & \ddots  & \ddots & \ddots  &  \\
\end{array}\right),
\end{eqnarray}
We note that for even $V(x)$, the quaternions of the outermost
band are such that

\begin{eqnarray}
\forall n,\qquad \zeta^{\bt}_{\pm d}(n)\equiv
\left[\begin{array}{cc} 0 &
0  \\
{(\zeta^{\bt}_{\pm d}(n))}_{10}     & 0  \\
\end{array}\right],{\hspace{1cm}}
\eta^{\bt}_{\pm d}(n)\equiv \left[\begin{array}{cc}
{(\eta^{\bt}_{\pm d}(n))}_{00} &
0  \\
{(\eta^{\bt}_{\pm d}(n))}_{10}     & {(\eta^{\bt}_{\pm d}(n))}_{11}  \\
\end{array}\right].
\end{eqnarray}

From (\ref{string}), we get the following quadratic relation
between the coefficients $({\zeta}^{\bt}_{j},{\eta}^{\bt}_{j})$
\begin{eqnarray}
\label{string1}
\sum_{j=l}^{d-1}{\eta}^{\bt}_{l-j}(n){\zeta}^{\bt}_{j}(n-l+j) -
\sum_{j=l}^{d-1}{\zeta}^{\bt}_{l-j}(n){\eta}^{\bt}_{j}(n-l+j)
={\zeta}^{\bt}_{l}(n).
\end{eqnarray}
This is the compatibility relation for the difference-differential
system and will be used in Section IX.

{\it Remark.} Here, one can notice the basic difference between
the properties of bi-orthogonal polynomials and their
skew-orthogonal counterparts. In the skew-orthogonal vectors, the
semi-infinite matrices are symmetric around the principal
diagonal. This is not true for the bi-orthogonal polynomials where
the matrices have an asymmetry.



\section{The Deformation Matrix}


In this section, we will consider infinitesimal deformation
corresponding to changes in $u_K$, the coefficient of the
potential $V(x)$. Using the definition of the semi-infinite
matrices $\Phi^{\bt}(x)$ and ${\tphi}^{\bt}(x)$ (and
$\Psi^{\bt}(x)$, ${\tpsi}^{\bt}(x)$), the deformation matrix can
be defined as:
\begin{eqnarray}
\label{d1} \frac{\partial}{\partial u_{K}}{\Phi}^{(\beta)}(x) &=&
U_{K}^{\bt}{\Phi}^{(\beta)}(x), \hspace{1cm}
 \frac{\partial}{\partial u_{K}}{\tphi}^{\bt}(x) = \tphi(x)
{[U^{\bt}_{K}]}^{D},{\hspace{1cm}}K=1,\ldots 2d, \\
\label{d2} \frac{\partial}{\partial u_{K}}{\Psi}^{(\beta)}(x) &=&
U_{K}^{\bt}{\Psi}^{(\beta)}(x), \hspace{1cm}
 \frac{\partial}{\partial u_{K}}{\tpsi}^{\bt}(x) = \tpsi(x)
{[U^{\bt}_{K}]}^{D},{\hspace{1cm}}K=1,\ldots 2d.
\end{eqnarray}
The matrix $U_K^{\bt}$ is anti-self dual, i.e.:
\begin{eqnarray}
\label{dual2}
 U_K^{\bt}=-{[U_K^{\bt}]}^D.
\end{eqnarray}

Moreover, they satisfy the following relations with $P$ and $R$
(we drop the superscript $\beta$ from the matrix $Q^{\bt}$,
$P^{\bt}$ and $R^{\bt}$ for simplicity):
\begin{eqnarray}
\label{string2}
\partial u_{K}P = [U_{K}^{\bt},P],\qquad
\partial u_{K}R = [U_{K}^{\bt},R].
\end{eqnarray}
Explicitly, $U_{K}^{\bt}$ can be written as
\begin{equation}
\label{U}
  U_{K}^{\bt}=-\frac{1}{K}\left(({{Q}^K)}_+
  - {({{{Q}}^K})_-^D}\right)
-\frac{1}{2K}{\left({({{Q}}^K)}_0 - {({{{Q}}^K})^D}_0\right)},
\end{equation}
where $Q_+$, $Q_-$ denote the upper and lower triangular
quaternion matrix while $Q_0$ is the diagonal quaternion.

{\bf Proof:}

Differentiating Eq.(\ref{ortho1}) w.r.t. $u_{K}$ and using Eqs.
(\ref{d1}) and (\ref{d2}), we get Eq.(\ref{dual2}).
 Eq.(\ref{string2}) follows by interchanging the operators $\partial/\partial x$ and
 $x\partial/\partial x$ with $\partial/\partial u_{K}$.
Finally to prove Eq.(\ref{U}), we start with  the $(2\times 1)$
normalized quasi-polynomial
\begin{equation}
{\Phi}^{(\beta)}_{n}(x):=\frac{1}{\sqrt{g^{(\beta)}_{n}}}e^{-V(x)}\Pi_n(x),\qquad
n\in\mathbb N,
\end{equation}
where $V(x)$ is defined in Eq.(\ref{V(x)}) and
\begin{eqnarray}
\Pi_n(x) = \left(\begin{array}{c}
\pi_{2n}(x)      \\
\pi_{2n+1}(x)   \\
\end{array}\right)
\end{eqnarray}
is the monic skew-orthogonal polynomial, defined in
Eq.(\ref{monopoly}).
 To save cluttering, we have suppressed the $\beta$ dependence of $\pi_{n}(x)$. We have used
$g_{{2n}}^{\bt}=g_{{2n+1}}^{\bt}$. Differentiating with respect to
the deformation parameter, we get
\begin{equation}
\label{deform}
\partial u_K{\Phi}^{(\beta)}_{n}(x)
=-\frac{1}{2g^{(\beta)}_{n}}\partial
u_{K}(g^{(\beta)}_{n}){\Phi}^{(\beta)}_{n}(x)
+\frac{e^{-V(x)}}{\sqrt {g^{(\beta)}_{n}}}\partial u_{K}\Pi_n(x)
-\frac{x^K}{K}{\Phi}^{(\beta)}_{n}(x).
\end{equation}
Also differentiating the skew-normalization condition, we get
\begin{eqnarray}
\nonumber -\frac{\partial}{\partial u_{K}}g^{(\beta)}_{n}
&=&-\int\left[\partial u_{K}{({\Phi}^{\bt}_n}(x))\tpsi_n^{\bt}(x)
+ {{\Phi}^{\bt}_n}(x)\partial u_{K}\tpsi_n^{\bt}(x)\right]dx,\\
&=&\frac{1}{K}{\left[Q^K+{(Q^K)}^{D}\right]}_{n,n}
{g^{(\beta)}_{n}}.
\end{eqnarray}

The diagonal elements have the form:
\begin{eqnarray}
\nonumber {(U_{K}^{\bt})}_{n,n} &=&
-{\left(\frac{Q^K}{K}\right)}_{n,n}
+\frac{1}{2K}{\left({({Q^K})}_{n,n}+{(Q^K)}^D_{n,n}
\right)},\\
&=& -\frac{1}{2K}\left[Q^K_{n,n}-{(Q^K)}^D_{n,n}\right].
\end{eqnarray}

Thus, we get
\begin{equation}
U_{K}^{\bt}=-\frac{1}{K}\left(Q^K_+ - {({Q^K})_-^D}\right)
-\frac{1}{2K}{\left({(Q^K_0)} - {({Q^K_0})}^D\right)}.
\end{equation}
{\it Remark.} Here the operation ${({Q^K})_-^D}$ means that the
dual is taken first and then the lower triangular part collected.
Also, due to the arbitrariness in the definition of an anti-self
dual matrix, we may choose the lower off-diagonal element of the
diagonal quaternions in $U_{K}^{\bt}$ zero. With this choice, we
can remove the arbitrariness in the definition of the odd
skew-orthonormal polynomials.



\section{The difference equation}
In this section, we introduce a sequence of companion-like
matrices ${A}_{N}^{\bt}(x)$ and $ {B}_{N}^{\bt}(x)$ of sizes
$(2d)\times (2d)$. Using the relations in Eqs. (\ref{beta4}) and
(\ref{beta1}), we can write:
\begin{eqnarray}
\label{difference1}
\sum_{k=-d}^{d}\alpha_k^{(4)}(n)\Phi^{(4)}_{n-k}(x)=0\qquad
\sum_{k=-d}^{d}\alpha_k^{(1)}(n)\Psi^{(1)}_{n-k}(x)=0\qquad\forall
k\in\mathbb N,
\end{eqnarray}
and
\begin{eqnarray}
\label{difference2}
\sum_{k=-d}^{d}\left(\alpha_k^{(1)}(n)-\delta_{0,k}\right)\Phi^{(1)}_{n-k}(x)=0\qquad
\sum_{k=-d}^{d}\left(\alpha_k^{(4)}(n)-\delta_{0,k}\right)\Psi^{(4)}_{n-k}(x)=0\qquad\forall
k\in\mathbb N,
\end{eqnarray}
we get for $N\geq d$:
\begin{eqnarray}
\label{recursion1}
 {\Phi}^{(4)}_{W+1}(x) &:=
& \left(\begin{array}{c}
\Phi^{(4)}_{N-d+1}(x)         \\
\vdots                      \\
\vdots                       \\
\Phi^{(4)}_{N+d}(x)                    \\
\end{array}\right)
= A_{N}^{(4)}(x) \left(\begin{array}{c}
\Phi^{(4)}_{N-d}(x)         \\
\vdots                      \\
\vdots                       \\
\Phi^{(4)}_{N+d-1}(x)                    \\
\end{array}\right)
={A_{N}^{(4)}}(x){\Phi}^{(4)}_{W}(x),
\end{eqnarray}
and
\begin{eqnarray}
\label{recursion1}
 {\Psi}^{(1)}_{W+1}(x) &:=
& \left(\begin{array}{c}
\Psi^{(1)}_{N-d+1}(x)         \\
\vdots                      \\
\vdots                       \\
\Psi^{(1)}_{N+d}(x)                    \\
\end{array}\right)
=A_{N}^{(1)}(x) \left(\begin{array}{c}
\Psi^{(1)}_{N-d}(x)         \\
\vdots                      \\
\vdots                       \\
\Psi^{(1)}_{N+d-1}(x)                    \\
\end{array}\right)
=A_{N}^{(1)}(x){\Psi}^{(1)}_{W}(x),
\end{eqnarray}
where (suppressing the $\beta$ dependence of $\alpha_j(n)$)
\begin{eqnarray}
{A_{N}^{\bt}}(x) &=& \left(\begin{array}{ccccc}
0    & {\bf 1}    & 0                   & \ldots            & 0 \\
0    & 0    &  {\bf 1}                        & \ddots            & 0        \\
\vdots    & \vdots                &  \vdots                  &  \vdots & {\bf 1}    \\
-{(\alpha_{-d}({{N}}))}^{-1}\alpha_{d}({{N}})
            &  \hdots        & -{(\alpha_{-d}({{N}}))}^{-1}(\alpha_{0}({N}))
             & \hdots
             & -{(\alpha_{-d}({{N}}))}^{-1}\alpha_{-d+1}({N})\\
\end{array}\right).
\end{eqnarray}
{\it Remark.} For potential $V(x)$ with even degree and  $\forall
n$, $\alpha_{\pm d}(n)=R^{\bt}_{n,n \pm d}-xP^{\bt}_{n,n \pm d}$
and its inverse has
 the quaternion structure:

\begin{eqnarray}
\label{abcd} {\alpha_{\pm d}({n})} &:= & \left(\begin{array}{cc}
a & 0        \\
b & c                   \\
\end{array}\right);{\hspace{1cm}}
{{\alpha_{\pm d}({n})}}^{-1} = \left(\begin{array}{cc}
\frac{1}{a} & 0        \\
-\frac{b}{ac} & \frac{1}{c}                   \\
\end{array}\right),{\hspace{1cm}} a,c\neq 0.
\end{eqnarray}
Hence they are invertible. This is the criteria for the existence
of the skew-orthogonal polynomials corresponding to even
potential. We leave it to the reader to verify that ${\alpha_{\pm
d}({n})}$ is non-invertible for potentials with odd degree.

 Using this, one can easily see that

\begin{eqnarray}
\label{recursion2} {\Phi^{(4)}_{W-1}}(x) &:= &
\left(\begin{array}{c}
\Phi^{(4)}_{N-d-1}(x)         \\
\vdots                      \\
\vdots                       \\
\Phi^{(4)}_{N+d-2}(x)                    \\
\end{array}\right)
= {(A_{N-1}^{(4)}}(x))^{-1}\left(\begin{array}{c}
\Phi^{(4)}_{N-d}(x)         \\
\vdots                      \\
\vdots                       \\
\Phi^{(4)}_{N+d-1}(x)                    \\
\end{array}\right)
={(A_{N-1}^{(4)}}(x))^{-1}{\Phi}^{(4)}_{W}(x),
\end{eqnarray}
and
\begin{eqnarray}
\label{recursion2} {\Psi^{(1)}_{W-1}}(x) &:= &
\left(\begin{array}{c}
\Psi^{(1)}_{N-d-1}(x)         \\
\vdots                      \\
\vdots                       \\
\Psi^{(1)}_{N+d-2}(x)                    \\
\end{array}\right)
= {(A_{N-1}^{(1)}}(x))^{-1}\left(\begin{array}{c}
\Psi^{(1)}_{N-d}(x)         \\
\vdots                      \\
\vdots                       \\
\Psi^{(1)}_{N+d-1}(x)                    \\
\end{array}\right)
={(A_{N-1}^{(1)}}(x))^{-1}{\Psi}^{(1)}_{W}(x),
\end{eqnarray}

 where
\begin{eqnarray}
 {(A_{N-1}^{\bt}}(x))^{-1} &=& \left(\begin{array}{ccccc}
-{(\alpha_{d}(N-1))}^{-1}\alpha_{d-1}(N-1)   & \hdots &
-{(\alpha_{d}(N-1))}^{-1}(\alpha_{0}(N-1))  & \hdots & -{(\alpha_{d}(N-1))}^{-1}\alpha_{-d}(N-1) \\
{\bf 1}         &    0               & \ldots & \ldots           & 0 \\
0                    &  \ddots       & \ddots & \ddots           & 0        \\
\vdots                    &  \vdots &  \vdots                 &  {\bf 1}   & 0   \\
\end{array}\right).
\end{eqnarray}
Similarly, for $\Psi^{(4)}(x)$ and $\Phi^{(1)}(x)$, $A^{\bt}(x)$
is replaced by $B^{\bt}(x)$, where
\begin{eqnarray}
{B_{N}^{\bt}}(x) &=& \left(\begin{array}{ccccc}
0    & {\bf 1}    & 0                   & \ldots            & 0 \\
0    & 0    &  {\bf 1}                        & \ddots            & 0        \\
\vdots    & \vdots                &  \vdots                  &  \vdots & {\bf 1}    \\
-{(\alpha_{-d}({{N}}))}^{-1}\alpha_{d}({{N}})
            &  \hdots        & -{(\alpha_{-d}({{N}}))}^{-1}(\alpha_{0}({N})-1)
             & \hdots
             & -{(\alpha_{-d}({{N}}))}^{-1}\alpha_{-d+1}({N})\\
\end{array}\right).
\end{eqnarray}
More generally, this ladder operator can be used successively to
obtain:

\begin{equation}
{\Phi^{\bt}_{W-j}}(x)={(A_{N-j}^{\bt}}(x))^{-1}\ldots
{(A_{N-1}^{\bt}}(x))^{-1} {\Phi^{\bt}_{W}}(x),\qquad\forall
j\in\mathbb N,
\end{equation}
and
\begin{equation}
{\Phi^{\bt}_{W+j}}(x)=A^{\bt}_{N+j-1}(x)\ldots A^{\bt}_{N}(x)
{\Phi^{\bt}_{W}}(x),\qquad\forall j\in\mathbb N.
\end{equation}
This is the underlying idea behind folding which will be discussed
in details in the next section.

\section{Folding}

{\it Remark}. From this point on we focus on the skew-orthogonal
vectors $\Phi_n^{(4)}(x)$ (and its dual $\Psi_n^{(1)}(x)$), but
everything being said can be immediately extended to the vectors
$\Phi_n^{(1)}(x)$, (and hence $\Psi_n^{(4)}(x)$) by interchanging
the r\^oles of matrices $R^{\bt}$ by $R^{\bt}-{\bf 1}$ .

The notion of ``folding'' is the following: we express any
quasipolynomial $\Phi_n^{\bt}(x)$ (or $\Psi_n^{\bt}(x)$) as a
linear combination of $2d$ {\em fixed} consecutive vectors
$\Phi_W^{\bt}(x)\equiv \Phi_{N-j}^{(4)}(x)\ ,\ j=d,\dots -d+1$
{\em with polynomial coefficients}. We now provide a way of
computing the folding function on the same line as done in
\cite{eynard3} for bi-orthogonal polynomials.

For a fixed subsequence (or window) $W$, such that $W\Rrightarrow
\{N-d,\ldots,N+d-1\}, N\geq d$, we seek to describe the folding of
the infinite wave-vector ${\Phi}^{\bt}(x)$ onto the window $W$ by
means of a single quaternion matrix $F(x)$ of size
$\infty\times(2d)$
 with polynomial entries such that:
\begin{equation} \forall n,\qquad \Phi_n^{\bt}(x) = \sum_{k=N-d}^{N+d-1}
{F}_{n,k}(x) \Phi_{k}^{\bt}(x)\ .
\end{equation}
(Similarly for $\Psi_n^{\bt}(x)$). In fact it is more convenient
to think  of $F(x)$ as a $\infty\times\infty$ matrix with only a
vertical band of width $2d$ of nonzero entries (with column index
in the range $N-d,\dots, N+d-1$).

Now, we will find an explicit form of the ``folding function'' in
terms of some upper and lower triangular matrices. We will derive
the folding function for $\Phi^{(4)}(x)$ and $\Psi^{(1)}(x)$. For
the folding function for $\Psi^{(4)}(x)$ and $\Phi^{(1)}(x)$ we
follow the same procedure and hence state the result without
proof. Note that everything will be written in terms of quaternion
matrices.

We will start with a few definitions and identities. We will
introduce the shift matrix $\Lambda$ comprising of $2\times 2$
blocks of unit matrix:
\begin{eqnarray}
\label{lambda}
 {\Lambda}= \left(\begin{array}{cccc}
0        &  {\bf 1}                  & 0            & \ldots \\
0                    &  0      & {\bf 1}            & 0        \\
0                    &  0                  & 0 & \ddots        \\
\vdots               &  \vdots        & \vdots            & \ddots        \\
\end{array}\right),
  \qquad {\mathbf 1}=\left(\begin{array}{cc}
1                  & 0   \\
0                  & 1\\
\end{array}\right),
\end{eqnarray}

satisfying the relation
\begin{equation}
{\Lambda}^t={\Lambda}^D,
\end{equation}
i.e., in this case, the transpose is the same as dual. Hence for
convenience, for the rest of the paper, we will denote it as
${\Lambda}^t$.

Let us define the projection matrices:
\begin{eqnarray}
\prod^{N} := {\bf 1}-\prod_{N},\qquad \prod^{N}_M :=
\prod_{N}-\prod_{M}.
\end{eqnarray}
We will also use the following identities,

\begin{equation}
\prod^{{N+d}}{(\Lambda^t)}^{d}={(\Lambda^t)}^{d}\prod^{N},\hspace{1cm}
\prod_{{N+d}}{(\Lambda^t)}^{d}={(\Lambda^t)}^{d}\prod_{N},{\hspace{1cm}}
\prod^{{N+d}}\prod_{{N+d}}=0,{\hspace{1cm}}
\prod^{{N+d}}\prod_{N}=0,
\end{equation}
\begin{equation}
\prod^{N}\prod_{{N+d}}=\prod_{{N+d}}-\prod_{N}=\prod_N^{N+d},
{\hspace{1cm}}
{({\Lambda}^t)}^d{\Lambda}^{d}=1-\prod_{d},{\hspace{1cm}}{\Lambda}^d\prod_d=0.
\end{equation}

To find an explicit formula for the matrix $F(x)$ we will use the
diagonal band-matrices $\alpha_{j}$, $j=-d,\ldots d$ to express
$R^{\bt}-xP^{\bt}$:
\begin{eqnarray} \label{p}{\overline R}^{\bt}(x)= R^{\bt}-xP^{\bt}  &=&
{\alpha}_{-d}{\Lambda}^{d}
       +\sum_{k=1}^{d-1}\left[\alpha_{-k}{\Lambda}^{k}\right]
       +\sum_{k=0}^{d}\left[\alpha_{k}{({\Lambda}^{t})}^{k}\right],
\end{eqnarray}
where $\Lambda$ is the shift matrix defined in Eq.(\ref{lambda}).
(For simplicity, we have again suppressed the $\beta$ dependence
of $\alpha_{j}$.)

Using the shift-matrix on Eq.(\ref{p}), we get
\begin{equation}
{(\Lambda^t)}^{d}\alpha_{-d}^{-1}[{\overline R}^{\bt}(x)]
={(\Lambda^t)}^{d}{(\Lambda)}^{d}
+{(\Lambda^t)}^{d}\alpha_{-d}^{-1}
\sum_{k=1}^{d-2}\left[\alpha_{-k}{\Lambda}^{k}\right]+
{(\Lambda^t)}^{d}\alpha_{-d}^{-1}
\sum_{k=0}^{d}\left[\alpha_{k}{({\Lambda}^{t})}^{k}\right].
\end{equation}

Thus we can define a semi-infinite matrix $G^{\bt}$, which is
strictly lower triangular:
\begin{equation}
\label{G} G^{\bt}:= 1- \prod_{d}-
{(\Lambda^t)}^{d}\alpha_{-d}^{-1}[{\overline R}^{\bt}(x)].
\end{equation}

Similarly

\begin{eqnarray}
\Lambda^{d}\alpha_{d}^{-1}[{\overline R}^{\bt}(x)] &=& 1
+\Lambda^{d}\alpha_{d}^{-1}\left[\sum_{k=1}^{d}\alpha_{-k}\Lambda^{k}
+\sum_{k=0}^{d-1}\alpha_{k}{(\Lambda^t)}^{k}\right],
\end{eqnarray}
allows us to define a strictly upper triangular matrix:
\begin{equation}
\label{C} C^{\bt}:=1-{\Lambda}^{d}{\alpha}_{d}^{-1}{\overline
R}^{\bt}(x).
\end{equation}

Multiplying Eq.(\ref{G}) by $\alpha_{-d}({\Lambda})^{d}$ on the
left and ${(1-G^{\bt})}^{-1}{(\Lambda^{t})}^{d}\alpha_{-d}^{-1}$
on the right, we get

\begin{equation}
{\bf 1}={\overline R}^{\bt}(x)
  \left[{(1-
  G^{\bt})}^{-1}{({\Lambda}^{t})}^{d}\alpha_{-d}^{-1}\right],
\end{equation}

while for the upper triangular matrix, we have the relation

\begin{equation}
{\bf 1}={(1-C^{\bt})}^{-1}\Lambda^{d}\alpha_{d}^{-1}{\overline
R}^{\bt}(x).
\end{equation}

Thus the matrix $[R^{\bt}-xP^{\bt}]$ has a left and right inverse,
but they are not the same. One is upper triangular while the other
is lower triangular.  However, they satisfy the following
relation:

\begin{equation}
{[{\overline R}^{\bt}(x)]}^{-1}_{L}=-{({[{\overline
R}^{\bt}(x)]}^{-1}_{R})}^{D}.
\end{equation}
One observes similar features for the matrix $(Q-x)$ in the
context of the two-matrix model studied in \cite{eynard3}.

Following the above procedure, we can also get the upper and lower
triangular matrix, needed to fold $\Psi^{(4)}(x)$ and
$\Phi^{(1)}(x)$. They are
\begin{equation}
\label{Btilde} {\tilde G}^{\bt}:= 1- \prod_{d}-
{(\Lambda^t)}^{d}\alpha_{-d}^{-1}[{\overline R}^{\bt}(x)-{\bf 1}],
{\hspace{1cm}}{\tilde
C}^{\bt}:=1-{\Lambda}^{d}{\alpha}_{d}^{-1}[{\overline
R}^{\bt}(x)-{\bf 1}],
\end{equation}
such that they satisfy
\begin{equation}
{\bf 1}=\left[[{\overline R}^{\bt}(x)-{\bf 1}]\right]
  \left[{(1- {\tilde
  G}^{\bt})}^{-1}{({\Lambda}^{t})}^{d}\alpha_{-d}^{-1}\right],
{\hspace{1cm}}{\bf 1}={(1-{\tilde
C}^{\bt})}^{-1}\Lambda^{d}\alpha_{d}^{-1}[{\overline
R}^{\bt}(x)-{\bf 1}].
\end{equation}


\subsection{Upper folding}

From the definition $[R^{(4)}-xP^{(4)}]\Phi^{(4)}(x)=0$, and Eqs.
(\ref{G}) and (\ref{C}), we get

\begin{equation}
\label{ab} \Phi^{(4)}(x) =C^{(4)}\Phi^{(4)}(x), \hspace{0.8cm}
\Phi^{(4)}(x) =\left(G^{(4)}+ \prod_{d+1} \right) \Phi^{(4)}(x).
\end{equation}
Similarly from the definition $[R^{(1)}-xP^{(1)}]\Psi^{(1)}(x)=0$,
and Eqs. (\ref{G}) and (\ref{C}), we get

\begin{equation}
\label{ab} \Psi^{(1)}(x) =C^{(1)}\Psi^{(1)}(x), \hspace{0.8cm}
\Psi^{(1)}(x) =\left(G^{(1)}+ \prod_{d+1} \right) \Psi^{(1)}(x).
\end{equation}

Given that the quaternion matrix $G$ is strictly lower triangular,
we can use Eq.(\ref{ab}) in an iterative manner to get

\begin{equation}
\nup \Phi^{(4)}(x) ={(1-G^{(4)})}^{-1} \nup G^{(4)} \ndown
\Phi^{(4)}(x),{\hspace{1cm}} \nup \Psi^{(1)}(x)
={(1-G^{(1)})}^{-1} \nup G^{(1)} \ndown \Psi^{(1)}(x).
\end{equation}
We will drop the parameter $\beta$ for simplification.
Also since the matrix $G$ has $d$ bands below the principal
diagonal, we have for the folding function:
\begin{equation}
\nbup F(x) ={(1-G)}^{-1} \nbup G \nbdown,
\end{equation}
where ${\overline N}=N+d-1$.
Simplifying further, we replace $G$ from Eq.(\ref{G}) to get
\begin{equation}
\nbup F(x) ={(1-G)}^{-1} \nbup \left[1-\prod_{d}-
\alpha_{-d}^{-1}{(\Lambda^t)}^{d}(R^{\bt}-xP^{\bt}) \right]
\nbdown.
\end{equation}
Then using the identities obtained before, one gets
\begin{eqnarray}
\nonumber \nbup F(x)
 &=& -{(1-G)}^{-1}
\nbup \alpha_{-d}^{-1}{({\Lambda^t})^{d}} [{\overline R}^{\bt}(x)]
\nbdown,
\\
\nonumber & = &
-\left[{(1-G)}^{-1}\alpha_{-d}^{-1}{({\Lambda^t})^{d}} \nup
({\overline R}^{\bt}(x)) \nbdown \right],
\\
\nonumber &=& -\prod^{\overline N}_{N}
-{(1-G)}^{-1}\alpha_{-d}^{-1}{({\Lambda^t})^{d}}\left[{\overline
R}^{\bt}(x),\prod_N\right].
\end{eqnarray}
Thus we have for the upper folding
\begin{equation}
\prod^{\overline N} F(x)=-\prod^{\overline N}_{N}
-{(1-G)}^{-1}\alpha_{-d}^{-1}{({\Lambda^t})^{d}}\left[{\overline
R}^{\bt}(x),\prod_N\right].
\end{equation}


\subsection{Lower Folding}

Now, we will study the lower folding function. We can write
\begin{equation}
\prod_M \Phi^{(4)}(x) ={(1-C^{(4)})}^{-1} \prod_M C^{(4)} \prod^M
\Phi^{(4)}(x), {\hspace{1cm}}\prod_M \Psi^{(1)}(x)
={(1-C^{(1)})}^{-1} \prod_M C^{(1)} \prod^M \Psi^{(1)}(x).
\end{equation}
where $M=N-d$. This relation is valid since we are folding finite
band matrices. Thus $F(x)$ can be written as
\begin{equation}
\prod_M F(x)= {(1-C)}^{-1} \prod_M C \prod^M.
\end{equation}
Replacing $C$ from Eq.(\ref{C}) and using the identities, we get

\begin{eqnarray}
\nonumber \prod_M F(x) &=& -{(1-C)}^{-1}{\Lambda}^{d}
\alpha_{d}^{-1}
 \prod_N
{\overline R}^{\bt}(x) \prod^M,
\\
\nonumber &=& -\prod_M^N +{(1-C)}^{-1}{\Lambda}^{d}
\alpha_{d}^{-1}[{\overline R}^{\bt}(x),\prod_N], \\
&=&
 -\prod^N_{M}
+{(1-C)}^{-1}{\Lambda}^{d}{\alpha}^{-1}_{d}[{\overline
R}^{\bt}(x),\prod_N].
\end{eqnarray}

\subsection{Folding Function}

The folding function is

\begin{eqnarray}
F(x)= \prod_M F(x) + \prod^{\overline N}F(x) + \prod^{N}_{M},
\end{eqnarray}
such that we get

\begin{equation}\label{mainformulaWn}
\encadremath{
\begin{array}{lll}
 F(x)=
\left({(1-C)}^{-1}{\Lambda}^{d}\alpha^{-1}_{d}
-{(1-G)}^{-1}{\alpha}^{-1}_{-d}{(\Lambda^t)}^{d}
\right)[{\overline R}^{\bt}(x),\prod_N].
\end{array}}
\end{equation}

One can easily repeat the above calculations to find the folding
function for $\Psi^{(4)}(x)$ and $\Phi^{(1)}(x)$. It gives:

\begin{equation}\label{mainformulaWn1} \encadremath{
\begin{array}{lll}
 {\tilde F(x)}=
\left({(1-{\tilde C})}^{-1}{\Lambda}^{d}\alpha^{-1}_{d}
-{(1-{\tilde G})}^{-1}{\alpha}^{-1}_{-d}{(\Lambda^t)}^{d}
\right)[{\overline R}^{\bt}(x),\prod_N].
\end{array}}
\end{equation}

{\it Remark.} In the context of bi-orthogonal polynomials, Bergere
and Eynard \cite{eynard5} have found a very elegant expression for
the folding function. It would be interesting to see the existence
of such compact expression in the context of skew-orthogonal
vectors.

\section{Folded Deformation Matrices}
 In this section, we will show that a finite sub-sequence or window of these
 skew-orthonormal vectors satisfy a system
 of PDE's under the infinitesimal change  of the deformation parameter. We will
 prove for $\Phi^{(4)}(x)$ and $\Psi^{(1)}(x)$. It can be trivially extended to $\Psi^{(4)}(x)$
 and $\Phi^{(1)}(x)$.
 We state the result.

 For convenience, in this section, we will drop the superscript $\beta$ from
 $Q^{\bt}$ as it has no importance on the deformation system. We define the
folded deformation matrix:
\begin{equation}
\mnb U_{K}^{\bt}F(x):={U^N_{K}}^{\bt},\qquad \mnb {
U}_{K}^{\bt}\tilde F(x):={\underline U^N_{K}}^{\bt},\qquad
K=1,\ldots 2d,
\end{equation}
where we project a $2d\times\infty$ quaternion matrix $\mnb
{U^{\bt}_{K}}$ onto a $\infty\times 2d$ quaternion matrix $F(x)$
and ${\tilde F}(x)$. We have
\begin{eqnarray}
\nonumber {-K{U^N_{K}}^{\bt}}
&=& \mnb
\left[[Q^K_{+}-(Q^K)^D_{-}]+\frac{1}{2}[Q^K_{0}-(Q^K_0)^D]\right]F(x),  \\
\nonumber
&=& \mnb
\left[[Q_{+}^K-(Q^K)^D_{-}]+\frac{1}{2}[Q^K_{0}-(Q^K_0)^D\right]
\left[ \mdown F(x) + \nbup F(x) + \mnb
\right],  \\
\nonumber
&=& -\mnb \left[KU^{\bt}_K\right] \mnb + \mnb Q^K \nbup F(x)- \mnb
(Q^K)^D \mdown F(x)
,  \\
\nonumber
&=& -\mnb \left[KU^{\bt}_K\right] \mnb - \mnb Q^K \left[
\nnb+{(1-G)}^{-1}{(\Lambda^t\alpha_{-d}^{-1})}^{d}\left[{\overline R}^{\bt}(x),\prod_N\right]\right] \\
\nonumber && - \mnb {(Q^K)}^D \left[ -\mn
+{(1-C)}^{-1}{(\alpha_{-d}^{-1}\Lambda)}^{d}\left[{\overline
R}^{\bt}(x),\prod_N]\right]
\right],  \\
\nonumber &=& -\mnb \left[KU^{\bt}_K\right] \mnb - \mnb Q^K \nnb +
\mnb {(Q^K)}^{D} \mn - \mnb \left[
(Q^K){(1-G)}^{-1}{(\Lambda^t\alpha_{-d}^{-1})}^{d}\left[{\overline
R}^{\bt}(x),\prod_N\right]
\right]  \\
\nonumber &&- \mnb \left[ ({Q^K}^D){(1-C)}^{-1}
{(\alpha_{-d}^{-1}\Lambda)}^{d}[P^{\bt},\prod_N]
\right],  \\
\nonumber &=& \mn \frac{{(Q_{0}^K+(Q_{0}^K)}^{D})}{2} \mn - \nbnn
\frac{(Q_{0}^K+{(Q_{0}^K)}^{D})}{2} \nbnn
-2x^K\left[\prod_{M}^{N}-\prod_{N}^{\overline N}\right]  \\
\nonumber && -\nbnn [Q^K-x^K]{({\overline
R}^{\bt}(x))}^{-1}_{R}\left[[{\overline
R}^{\bt}(x),\prod_N]\right] - \mnb {(Q^K-x^K)}^{D}{({\overline
R}^{\bt}(x))}^{-1}_{L}\left[{\overline R}^{\bt}(x),\prod_N\right]
,  \\
&=& \prod_{M}^{\overline N}\frac{{(Q_{0}^K+(Q_{0}^K)}^{D})}{2}
\left[\prod_{M}^{N}-\prod_{N}^{\overline N}\right] - \mnb {\cal
W}_{K}(x)\left[[{\overline R}^{\bt}(x),\prod_N]\right]
-2x^K\left[\prod_{M}^{N}-\prod_{N}^{\overline N}\right],
\end{eqnarray}
where
\begin{eqnarray}
{\cal W}_{K}(x)&=& {(Q^K-x^K)}{({\overline R}^{\bt}(x))}^{-1}_{R}
             +{(Q^K-x^K)}^{D}{({\overline R}^{\bt}(x))}^{-1}_{L}.
\end{eqnarray}
Now, if we define in analogy to the Pauli matrix $\sigma_3$,
\begin{equation}
\Sigma_{3}:=\left[\prod_{M}^{N}-\prod_{N}^{\overline
N}\right]={\rm diag}(\underbrace{{\mathbf 1}\ldots{\mathbf
1}}_{d},\underbrace{{\mathbf {-1}}\ldots{\mathbf {-1}}}_{d}),
\end{equation}
then we may write
\begin{equation}\label{mainformulaWn3} \encadremath{
\begin{array}{lll}
{-K{U^N_{K}}^{\bt}} = \prod_{M}^{\overline
N}\frac{{(Q_{0}^K+(Q_{0}^K)}^{D})}{2} \Sigma_{3}- \mnb {\cal
W}_{K}(x)\left[{\overline R}^{\bt}(x),\prod_N\right] -2x^K
\Sigma_{3},
\end{array}}
\end{equation}
For $\Psi^{(4)}(x)$ and $\Phi^{(1)}(x)$, we have

\begin{equation}\label{mainformulaWn4} \encadremath{
\begin{array}{lll}
{-K{\underline U^N_{K}}^{\bt}} &=& \prod_{M}^{\overline
N}\frac{{(Q_{0}^K+(Q_{0}^K)}^{D})}{2} \Sigma_{3} - \mnb {\cal
\underline W}_{K}(x)\left[{\overline R}^{\bt}(x),\prod_N\right]
-2x^K\Sigma_{3},
\end{array}}
\end{equation}
where
\begin{eqnarray}
\underline {\cal W}_{K}(x)&=& {(Q^K-x^K)}{({\overline
R}^{\bt}(x)-{\bf 1})}^{-1}_{R}
             +{(Q^K-x^K)}^{D}{({\overline R}^{\bt}(x)-{\bf 1})}^{-1}_{L}.
\end{eqnarray}
 Thus, we see that these skew-orthogonal vectors satisfy
a system of PDE's with respect to the deformation parameter
 and is given by
\begin{eqnarray}
\frac{\partial}{\partial u_K}{\Phi}_{W}^{(4)}(x) &=&
{U_K^N}^{(4)}{\Phi}_{W}^{(4)}(x),{\hspace{1cm}}
\frac{\partial}{\partial u_K}{\Psi}_{W}^{{(1)}}(x)
={U_K^N}^{(1)}{\Psi}_{W}^{{(1)}}(x),\\
\frac{\partial}{\partial u_K}{\Psi}_{W}^{(4)}(x) &=&
{\underline
U_K^N}^{(4)}{\Psi}_{W}^{(4)}(x),{\hspace{1cm}}\frac{\partial}{\partial
u_K}{\Phi}_{W}^{{(1)}}(x)=
{\underline U_K^N}^{(1)}{\Phi}_{W}^{{(1)}}(x),
\end{eqnarray}
where ${\Phi}^{\bt}_{W}(x)$ and ${\Psi}^{\bt}_{W}(x)$is defined in
Eqs.(\ref{window},\ref{window1}).


\section{The differential equation}


Finally, using the results of the deformation equation, we will
derive the ODEs satisfied by the finite subsequence of
skew-orthogonal vectors.

{\it Remark.} We will calculate explicitly the differential
operator for $\Phi^{(4)}(x)$ and $\Psi^{(4)}(x)$. The
corresponding dual vectors $\Psi^{(1)}(x)$ and $\Phi^{(1)}(x)$
respectively can be obtained by a simple transfer of
$Q^{(4)}\mapsto Q^{(1)}$ and hence $P^{(4)}\mapsto P^{(1)}$ and
$R^{(4)}\mapsto R^{(1)}$. Everything else remains the same.

We recall that the differential operator on any of these vectors:
\begin{eqnarray}
\nonumber \frac{d}{dx}\Phi^{\bt}(x) &=& P^{\bt}\Phi^{\bt}(x)=
-[(V'(Q))_{+}-{(V'(Q))}^{D}_{-}+\frac{1}{2}[{(V'(Q))}_{0}-{{(V'(Q))}}^D_{0}]]\Phi^{\bt}(x)\\
&=&
-\sum_{K=0}^{2d-1}u_{K+1}\left[Q^{K}_{+}-{(Q^{K})}^{D}_{-}+\frac{1}{2}[Q^{K}_{0}-(Q^{K})^D_{0}\right]\Phi^{\bt}(x),
\end{eqnarray}
and similarly for $\Psi^{\bt}(x)$. Since ${D_{N}^{\bt}}$ and
${{\underline D}_{N}^{\bt}}$ is the folded version of $P^{\bt}$,
we observe that the differential equation satisfied by the finite
subsequence of $\Phi^{(4)}(x)$, namely ${\Phi}^{(4)}_{W}(x)$, is
given by:

\begin{eqnarray}
\nonumber\frac{d}{dx}{\Phi}^{(4)}_{W}(x)
&=& \mnb P^{(4)}F(x)\Phi^{(4)}(x),\\
&=& {D_{N}^{(4)}}\Phi^{(4)}_{W}(x),
\end{eqnarray}
such that with $Q:=Q^{(4)}$, $P:=P^{(4)}$ and $R:=R^{(4)}$, we
have
\begin{eqnarray}
\nonumber
 {D_{N}^{(4)}}=\mnb PF(x)&=&
-\mnb [(V'(Q))_{+}-{(V'(Q))}^{D}_{-}+\frac{1}{2}[{(V'(Q))}_{0}-{{(V'(Q))}}^D_{0}]]F(x),\\
\nonumber   &=&
-\sum_{K=0}^{2d-1}u_{K+1}\mnb\left[Q^{K}_{+}-{(Q^{K})}^{D}_{-}+\frac{1}{2}[Q^{K}_{0}-(Q^{K})^D_{0}\right]F(x),\\
\nonumber
  &=& -\sum_{K=0}^{2d-1}u_{K+1}({-KU_{K}^N}^{(4)}),\\
  &=& 2V'(x)\Sigma_{3}+\mnb{\cal W}_{K}^N(x)\left[{\overline R}(x),\prod_{N}\right].
\end{eqnarray}
In the last step, we have used the anti-self dual property of $P$.

Thus we have
\begin{equation}\label{mainformulaWn5} \encadremath{
\begin{array}{lll}
\frac{d}{dx}{\Phi}^{(4)}_{W}(x)&=& \mnb PF(x)\Phi^{(4)}(x)
  = \left[2V'(x)\Sigma_{3}+\mnb{\cal W}_{K}^N(x)\left[{\overline R}(x),\prod_{N}\right]\right]{\Phi}^{(4)}_{W}(x).
\end{array}}
\end{equation}

 where

\begin{eqnarray}
{\cal W_{K}^N}(x) &=&
 [V'(Q)-V'(x)]{{(xP-R)}^{-1}_{R}}+{[(V'(Q))-V'(x)]}^{D}{{(xP-R)}^{-1}_{L}}.
\end{eqnarray}
Replacing $\Phi^{(4)}(x)\mapsto\Psi^{(1)}(x)$ and $Q^{(4)}\mapsto
Q^{(1)}$, (and hence $P^{(4)}\mapsto P^{(1)}$ and $R^{(4)}\mapsto
R^{(1)}$ ) we can obtain the differential equation for the dual
vector $\Psi^{(1)}(x)$.

Explicitly, we have
\begin{eqnarray}
\nonumber {D_{N}^{(4)}} &=&
 2V'(x)\left(\begin{array}{cccccc}
{\bf 1}    & 0                 & 0                   & \ldots            &0&0    \\
0          & \ddots            & 0                   & 0                 &0&0    \\
0          & 0                 & {\bf 1}             & \ldots            &0&0    \\
0          & 0                 & 0                   & -{\bf 1}                &0&0    \\
0          & 0                 & 0                   & 0            &\ddots & 0    \\
0          & 0                 & 0                   & 0                &0& -{\bf 1}    \\
\end{array}\right)_{2d\times 2d} +\\ && {\cal
W_{K}^N}(x) \left(\begin{array}{ccccccc}
0  &    0  &  0  &{\alpha}_{-d}(M)  &  0                 &  0 \\
0  &    0  &  0  &\vdots            &\ddots              &  0 \\
0  &    0  &  0  &{\alpha}_{-1}(N)  &\ldots              &{\alpha}_{-d}(N)\\
-{\alpha}_{d}(N+1)   &   \ldots     &-{\alpha}_{1}(N+1)  &  0  &  0    &  0\\
0& \ddots& \vdots &  0                  &0&0 \\
0&0&-{\alpha}_{d}({\overline N})                    &0&0&0 \\
\end{array}\right).
\end{eqnarray}

For $\Psi_W^{(4)}(x)$  we have
\begin{equation}\label{mainformulaWn6} \encadremath{
\begin{array}{lll}
\frac{d}{dx}{\Psi}^{(4)}_{W}(x) &=&\mnb P{\tilde
F}(x)\Psi^{(4)}(x)
 = \left[2V'(x)\Sigma_{3}+\mnb{\cal\underline W}_{K}^N(x)
 \left[{\overline R}(x),\prod_{N}\right]\right]{\Psi}^{(4)}_{W}(x),
\end{array}}
\end{equation}

where
 ${\cal\underline W_{K}^N}(x)$ is given by
\begin{eqnarray}
{\cal\underline W_{K}^N}(x) &=&
 [V'(Q)-V'(x)]{{({\overline
 R}(x)-1)}^{-1}_{R}}+{[(V'(Q))-V'(x)]}^{D}{{({\overline
 R}(x)-1)}^{-1}_{L}}.
\end{eqnarray}
For the dual vector $\Phi_W^{(1)}(x)$, we follow the same
procedure. Replacing $\Psi^{(4)}(x)\mapsto\Phi^{(1)}(x)$ and
$Q^{(4)}\mapsto Q^{(1)}$, (and hence $P^{(4)}\mapsto P^{(1)}$ and
$R^{(4)}\mapsto R^{(1)}$ ) we can obtain the ODE's  for the dual
vector $\Phi^{(1)}(x)$.

\section{Fundamental solutions}
We have seen that a finite sub-sequence of $(2\times 1)$
skew-orthogonal vectors $\Phi_{N-k}^{\bt}(x)$ and
$\Psi_{N-k}^{\bt}(x)$, $k=-d+1,\ldots d$, satisfy a system of
differential-difference-deformation equation. To obtain the
fundamental solutions, useful in the Riemann-Hilbert analysis, we
need to look for $2d$ other solutions. In this section, we will
show that the Cauchy-like transforms of these skew-orthogonal
vectors indeed form such solutions.

We will look for some integral representations of these
skew-orthogonal vectors as the possible ``other solutions''. From
the previous experience in orthogonal \cite{bleher} and
bi-orthogonal polynomials \cite{eynard4}, we look for some
Cauchy-like transforms ${\tilde\Psi}^{\bt}_n(x)$ and
${\tilde\Phi}^{\bt}_n(x)$, and hope that at least for sufficiently
large $n$, these functions will form the simultaneous solutions
for the system of difference-differential-deformation equations.
The remaining solutions can be obtained by taking the Cauchy-like
transforms of certain moment functions $f_j(x)$ of order $j$, to
be defined below. This trick had been used previously to define
Riemann-Hilbert problems for bi-orthogonal
 \cite{mclaughlin} and skew-orthogonal polynomials \cite{pierce} respectively.

Let us define

\begin{eqnarray}
\label{fj} {f}_j(x) := \exp[V(x)]\int_{\mathbb R}\epsilon(x-y)
y^j\exp[-V(y)]dy,\qquad j=0,\ldots,2d-2.
\end{eqnarray}

We will show that the functions:

\begin{eqnarray}
{\Psi}^{(j,\beta)}_n(x) & := & \int_{\mathbb
R}\frac{f_j(x)-f_j(z)}{x-z}{\Psi}^{(\beta)}_n(z)\exp[-V(z)]dz,\qquad j=0,\ldots, 2d-2,\\
{\Phi}^{(j,\beta)}_n(x) & := & \int_{\mathbb
R}\frac{f_j(x)-f_j(z)}{x-z}{\Phi}^{(\beta)}_n(z)\exp[-V(z)]dz\qquad
j=0,\ldots, 2d-2,
\end{eqnarray}
and the Cauchy-like transform of the skew-orthogonal vectors:
\begin{eqnarray}
{\tilde\Psi}^{\bt}_n(x) = \exp[V(x)]\int_{\mathbb
R}\frac{{\Psi}^{\bt}_{n}(z)\exp[-V(z)]}{(x-z)}dz,
\qquad{\tilde\Phi}^{\bt}_n(x) = \exp[V(x)]\int_{\mathbb
R}\frac{{\Phi}^{\bt}_{n}(z)\exp[-V(z)]}{(x-z)}dz,\qquad
n\in\mathbb N,
\end{eqnarray}
are simultaneous solutions of the
differential-difference-deformation equations.

\subsection{The Cauchy-like transform}
Componentwise, we get using Eqs. (\ref{PR4}) and (\ref{PR1}):

\begin{eqnarray}
\label{psiphitilde1} {\tilde\psi}^{(4)}_n(x) = \sum
P^{(4)}_{n,m}{\tilde\phi}^{(4)}_m(x),
\qquad{\tilde\phi}^{(1)}_n(x) = \sum
P^{(1)}_{n,m}{\tilde\psi}^{(1)}_m(x).
\end{eqnarray}

We also get
\begin{eqnarray}
\nonumber \frac{d}{dx}{\tilde\psi}^{\bt}_{n}(x) &=&
\exp[V(x)]\int_{\mathbb
R}\frac{V'(x)-V'(z)}{x-z}{\psi}^{\bt}_n(z)\exp[-V(z)]dz+\sum
P^{\bt}_{n,m}{\tilde\psi}^{\bt}_m(x),\\
&=& \exp[V(x)]\sum_{m=0}^{2d-2}C_{m}^{\bt}(x)Z_{m,n}+\sum
P^{\bt}_{n,m}{\tilde\psi}^{\bt}_m(x),
\end{eqnarray}
where $V'(x)=\frac{d}{dx}V(x)$ and $V'(z)=\frac{d}{dz}V(z)$. Here,
we have used
$\exp[-V(z)]\left[\{V'(x)-V'(z)\}/(x-z)\right]=\sum_{m=0}^{2d-2}C_{m}^{\bt}(x)\phi_{m}^{\bt}(z)$.
Thus we have
\begin{eqnarray}
\label{ptilde1} \frac{d}{dx}{\tilde\psi}^{\bt}_{n}(x) = \sum
P^{\bt}_{n,m}{\tilde\psi}^{\bt}_m(x),\qquad n\geq 2d.
\end{eqnarray}

Using Eqs.(\ref{psiphitilde1}) and (\ref{ptilde1}), we get
\begin{eqnarray}
\label{phitildeprime1} {{\tilde\phi}}^{\prime(4)}_n(x) =
{\tilde\psi}^{(4)}_n(x)=\sum P^{(4)}_{n,m}{\tilde\phi}^{(4)}_m(x),
\qquad{\tilde\psi}^{\prime(1)}_n(x) =
2{\tilde\phi}^{(1)}_n(x),\qquad {{\tilde\phi}}^{\prime(1)}_n(x) =
\sum P^{(1)}_{n,m}{\tilde\phi}^{(1)}_m(x),\qquad n\geq 2d.
\end{eqnarray}
Thus for $n\geq 2d-1$, our chosen functions  satisfy the same
system of differential equations as done by the skew-orthogonal
vectors $\phi_n^{\bt}(x)$ and $\psi_n^{\bt}(x)$ (\ref{recursion}).

Having obtained the relations between the ${\tilde\phi}^{\bt}(x)$
and ${\tilde\psi}^{\bt}(x)$, we will now show that these functions
also satisfy the same recursion relations. We start with
\begin{eqnarray}\label{rtilde} \nonumber
x\frac{d}{dx}{\tilde\psi}^{\bt}_n(x) &=& x\frac{d}{dx}
\exp[V(x)]\int_{\mathbb R}\frac{{\psi}^{\bt}_{n}(z)\exp[-V(z)]}{(x-z)}dz,\\
\nonumber &=& xV'(x){\tilde\psi}^{\bt}_{n}(x)
-\exp[V(x)]\int_{\mathbb
R}\frac{xV'(z)}{x-z}\psi_{n}^{\bt}(z)\exp[-V(z)]dz
+\exp[V(x)]\int_{\mathbb
R}\frac{x{\psi_{n}^{\bt}}'(z)}{x-z}\exp[-V(z)]dz.
\end{eqnarray}
Now, replacing $x\rightarrow x-z+z$ on the right hand side, we get
\begin{eqnarray}
\nonumber x\frac{d}{dx}{\tilde\psi}^{\bt}_{n}(x) &=&
\exp[V(x)]\int_{\mathbb
R}\frac{xV'(x)-zV'(z)}{x-z}\psi_{n}^{\bt}(z)\exp[-V(z)]dz
+\exp[V(x)]\int_{\mathbb R}\frac{d}{dz}\left(\psi_{n}^{\bt}(z)\exp[-V(z)]\right)dz\\
&& +\exp[V(x)]\int_{\mathbb
R}\frac{\exp[-V(z)]}{x-z}\left[z\frac{d}{dz}(\psi_{n}^{\bt}(z)\right]dz.
\end{eqnarray}
Thus we have

\begin{eqnarray}
\label{ddxtildepsi} \nonumber x\frac{d}{dx}{\tilde\psi}^{\bt}_n(x)
&=& \exp[V(x)]\sum_{m=0}^{2d-1}d_{m}^{\bt}(x)Z_{m,n}+\sum
R^{\bt}_{nm}{\tilde\psi}^{\bt}_n(x),\hspace{2.5cm} \beta=1,\\
&=& \exp[V(x)]\sum_{m=0}^{2d-1}d_{m}^{\bt}(x)Z_{m,n}+\sum
(R^{\bt}_{nm}-\delta_{n,m}){\tilde\psi}^{\bt}_n(x),\hspace{1cm}
\beta=4.
\end{eqnarray}
Here, we have used
$\exp[-V(z)]\left[\{xV'(x)-zV'(z)\}/(x-z)\right]=\sum_{m=0}^{2d-1}d_{m}^{\bt}(x)\phi_{m}^{\bt}(z)$.
Thus for $n \geq 2d$, we can see that the first term drops out and
hence ${\tilde\psi}^{\bt}_{n}(x)$ satisfy Eqs.(\ref{recursion4},
\ref{recursion5}) respectively.

Similarly, for $\beta=4$, we get for $n\geq 2d$,
\begin{eqnarray}
\label{beta4tilde} \nonumber x\frac{d}{dx}{\tilde\phi}^{(4)}_n(x)
&=& x{\tilde\psi}^{(4)}_n(x), \\
\nonumber &=&
\exp[V(x)]\int_{\mathbb R}\frac{(x-z+z)}{x-z}\psi_{n}^{(4)}(z)\exp[-V(z)]dz,\\
&=& \sum R^{(4)}_{n,m}{\tilde\phi}^{(4)}_m(x),
\end{eqnarray}
where the first term drops off due to the condition $n\geq 2d$,
$d\geq 1$. For $\beta=1$, we differentiate (\ref{ddxtildepsi}) to
get
\begin{eqnarray}
x\frac{d}{dx}{\tilde\phi}^{(1)}_n(x) &=& \sum
(R^{(1)}_{nm}-\delta_{n,m}){\tilde\phi}^{(1)}_n(x),\qquad n\geq
2d.
\end{eqnarray}
This in turn proves that for $n\geq 2d$, the Cauchy-like
transforms satisfy the same set of recursion relations
(\ref{difference1}) and (\ref{difference2}) as done by the
skew-orthogonal vectors.

Thus we prove that our function is compatible under a system of
difference-differential equation, i.e. the solution of the system
of differential equation is also a simultaneous solution of the
system of difference equation. We will now show that these
solutions are also compatible with the system of deformation
relations i.e. they are also solutions of the system of PDE's.

We can see that
\begin{eqnarray}
\label{psitildeu1} \nonumber \frac{\partial}{\partial
u_K}{\tilde\psi}^{\bt}_n(x) &=& \frac{\exp[V(x)]}{K}\int_{\mathbb
R}\frac{x^K-z^K}{x-z}\psi^{\bt}_{n}(z)\exp[-V(z)]dz
+\sum {(U_K^{\beta})}_{nm}{\tilde\psi}^{\bt}_m(x),\\
&=& \exp[V(x)]\sum_{m=0}^{k-1}a_{m}^{\bt}(x)Z_{m,n}+\sum
{(U_K^{\beta})}_{n,m}{\tilde\psi}^{\bt}_m(x).
\end{eqnarray}
where
$\exp[-V(z)]\left[\{x^K-z^K\}/(x-z)\right]=\sum_{m=0}^{K-1}a_{m}^{\bt}(x)\phi_{m}^{\bt}(z)$.

Thus for $n\geq 2d$, ${\tilde\psi}^{\bt}_n(x)$ satisfy
Eq.(\ref{d2}). Also, combining Eqs.(\ref{phitildeprime1},
\ref{psitildeu1}), we can see that

\begin{eqnarray}
{{\tilde\phi}}^{\bt}_n(x) &=&\sum
{(U_K^{\beta})}_{n,m}{\tilde\phi}^{\bt}_m(x),\qquad n\geq 2d.
\end{eqnarray}

Thus for $n \geq 2d$, the new-functions ${\tilde\phi}_n^{\bt}(x)$
and ${\tilde\psi}_n^{\bt}(x)$ satisfy the system of
differential-difference-deformation equation and hence are
mutually compatible.

\subsection{The other solutions}

%

In this subsection, we will show that ${\Psi}^{(j,\beta)}_n(x)$
and ${\Phi}^{(j,\beta)}_n(x)$, $j=0,\ldots, 2d-2$, are also
solutions to the system of difference-deformation-differential
equations. For this, we will characterize the skew-orthogonal
functions $\phi^{(1)}_{k}(x)$ and $\psi^{(4)}_{k}(x)$ (which are
polynomials of degree $k$ and $k+2d-1$ respectively) through a set
of skew-orthogonality relations with respect to the functions
$f_{j}(x)$. The expressions for these auxiliary solutions were
suggested (for the case $\beta=1$) by the common referee of this
paper and \cite{pierce} using expressions contained therein, and
extended by the present author to the case $\beta=4$.

We will start with the identity for a polynomial $\pi_j(x)$ of
order $j$ as:

\begin{eqnarray}
\label{pij} \pi_j(x)=\exp[V(x)]\frac{d}{dx}(x^{j-2d+1}\exp[-V(x)])
\end{eqnarray}
There are $2k+1$ orthogonality property for the skew-orthogonal
polynomials of order $2k$ and $2k+1$. Starting with
\begin{eqnarray}
Z_{j+2d-1,2k} &=&\int_{\mathbb
R}\psi^{\bt}_{2k}\pi_{j+2d-1}\exp[-V(x)]dx=0,\qquad \forall
j=0,\ldots,2k-2d+1,
\end{eqnarray}
and using Eq.(\ref{pij}), we get the $2k-2d+2$ conditions for
\begin{eqnarray}
\beta=1,\qquad Z_{2k,j+2d-1}=2\int_{\mathbb
R}\phi^{(1)}_{2k}(x)x^j\exp[-V(x)]dx=0,\qquad \forall
j=0,\ldots,2k-2d+1,
\end{eqnarray}
and
\begin{eqnarray}
\beta=4,\qquad Z_{2k,j+2d-1}=\int_{\mathbb
R}\psi^{'(4)}_{2k}(x)x^j\exp[-V(x)]dx=0,\qquad \forall
j=0,\ldots,2k-2d+1,
\end{eqnarray}
while for the remaining $2d-1$ conditions, one can have

\begin{eqnarray}
Z_{2k,j}=\int_{\mathbb R}\int_{\mathbb
R}\phi_{2k}^{(1)}(x)y^{j}\epsilon(x-y)\exp[-V(y)]dxdy &=& 0,\qquad
j=0,\ldots,2d-2,
\end{eqnarray}
\begin{eqnarray}
Z_{j,2k}=\int_{\mathbb R}\int_{\mathbb
R}\psi_{2k}^{(4)}(x)y^{j+2d-1}\epsilon(x-y)\exp[-V(y)]dxdy &=&
0,\qquad j=0,\ldots,2d-2.
\end{eqnarray}

Thus we define the function $f_{j}(x)$:
\begin{eqnarray}
\label{fj} {f}_j(x) := \exp[V(x)]\int_{\mathbb R}\epsilon(x-y)
y^j\exp[-V(y)]dy,\qquad j=0,\ldots,2d-2.
\end{eqnarray}
{\it Remark}: Here, we note that for $\beta=4$, without loss of
generality, $f_{j}(x)$ can also be defined as:
\begin{eqnarray}
\label{fj} {f}_j(x) := \exp[V(x)]\int y^j\exp[-V(y)]dy+c,\qquad
j=0,\ldots,2d-2,
\end{eqnarray}
where $c$ is a constant that can be fixed by
skew-orthonormalization condition.

Having defined $f_j(x)$, we will show that the functions:

\begin{eqnarray}
{\Psi}^{(j,\beta)}_n(x) := \int_{\mathbb
R}\frac{f_j(x)-f_j(z)}{x-z}{\Psi}^{(\beta)}_n(z)\exp[-V(z)]dz;
\qquad {\Phi}^{(j,\beta)}_n(x) := \int_{\mathbb
R}\frac{f_j(x)-f_j(z)}{x-z}{\Phi}^{(\beta)}_n(z)\exp[-V(z)]dz,
\end{eqnarray}
satisfy the same system of difference-differential-deformation
equations.

Componentwise, using Eqs. (\ref{PR4}) and (\ref{PR1}), we get

\begin{eqnarray}
\label{psiphitilde} {\psi}^{(j,4)}_n(x) = \sum
P^{(4)}_{n,m}{\phi}^{(j,4)}_m(x), \qquad 2{\phi}^{(j,1)}_n(x) =
\sum P^{(1)}_{n,m}{\psi}^{(j,1)}_m(x).
\end{eqnarray}

We also get
\begin{eqnarray}
\nonumber \frac{d}{dx}{\psi}^{(j,\beta)}_{n}(x) &=&
f_j(x)\int_{\mathbb
R}\frac{V'(x)-V'(z)}{x-z}{\psi}^{\bt}_n(z)\exp[-V(z)]dz
+2\int_{\mathbb R}\frac{x^j-z^j}{x-z}{\psi}^{\bt}_n(z)\exp[-V(z)]dz\\
\nonumber &&+\int_{\mathbb R}\frac{f_j(x)-f_j(z)}{x-z}{\psi
'}^{\bt}_n(z)\exp[-V(z)]dz\\
&=& f_j(x)\sum_{m=0}^{2d-2}C_{m}^{\bt}(x)Z_{m,n}+
2\sum_{m=0}^{j-1}a_{m}^{\bt}(x)Z_{m,n}+\sum
P^{\bt}_{n,m}{\psi}^{(j,\beta)}_m(x).
\end{eqnarray}
Thus we have
\begin{eqnarray}
\label{ptilde} \frac{d}{dx}{\psi}^{(j,\beta)}_{n}(x) = \sum
P^{\bt}_{n,m}{\psi}^{(j,\beta)}_m(x),\qquad n\geq 2d.
\end{eqnarray}

Using Eqs.(\ref{psiphitilde}) and (\ref{ptilde}), we get
\begin{eqnarray}
\label{phitildeprime} {{\phi}}^{\prime(j,4)}_n(x) =
{\psi}^{(j,4)}_n(x)=\sum P^{(4)}_{n,m}{\phi}^{(j,4)}_m(x),
\qquad{\psi}^{\prime(j,1)}_n(x) = 2{\phi}^{(j,1)}_n(x),\qquad
{{\phi}}^{\prime(j,1)}_n(x) = \sum
P^{(1)}_{n,m}{\phi}^{(j,1)}_m(x),\qquad n\geq 2d.
\end{eqnarray}
Thus for $n\geq 2d$, our chosen functions  satisfy the same system
of differential equations as done by the skew-orthogonal vectors
$\phi_n^{\bt}(x)$ and $\psi_n^{\bt}(x)$ (\ref{recursion}).

Having obtained the relations between the ${\phi}^{(j,\beta)}(x)$
and ${\psi}^{(j,\beta)}(x)$, we will now show that these functions
also satisfy the same recursion relations. We start with
\begin{eqnarray}\label{rtilde} \nonumber
x\frac{d}{dx}{\psi}^{(j,\beta)}_n(x) &=& f_j(x) \int_{\mathbb
R}\frac{(xV'(x)-zV'(z))}{(x-z)}{\psi}^{\bt}_{n}(z)\exp[-V(z)]dz
+2\int_{\mathbb R}\frac{(x^{j+1}-z^{j+1})}{(x-z)}{\psi}^{\bt}_{n}(z)\exp[-V(z)]dz\\
&& +\int_{\mathbb
R}\frac{d}{dz}\left({\psi}^{\bt}_{n}(z)(f_j(x)-f_j(z))\exp[-V(z)]\right)dz
+ \int_{\mathbb
R}z{\psi'}^{\bt}_{n}(z)\frac{f_j(x)-f_j(z)}{x-z}\exp[-V(z)]dz ,
\end{eqnarray}
which for $\beta=1$ gives
\begin{eqnarray}
x\frac{d}{dx}{\psi}^{(j,1)}_n(x)&=&
f_j(x)\sum_{m=0}^{2d-1}d_{m}^{(1)}(x)Z_{m,n}+
2\sum_{m=0}^{j}a_{m}^{(1)}(x)Z_{m,n} +\sum
R^{(1)}_{nm}{\psi}^{(j,1)}_n(x),\qquad j=0,\ldots,2d-2,
\end{eqnarray}
and for $\beta=4$ gives
\begin{eqnarray}
x\frac{d}{dx}{\psi}^{(j,4)}_n(x)&=&
f_j(x)\sum_{m=0}^{2d-1}d_{m}^{(4)}(x)Z_{m,n}+
2\sum_{m=0}^{j}a_{m}^{(4)}(x)Z_{m,n} +\sum
(R^{(4)}_{nm}-\delta_{n,m}){\psi}^{(j,4)}_n(x),\qquad
j=0,\ldots,2d-2.
\end{eqnarray}
Thus for $n \geq 2d$, we can see that the first two terms drop out
and hence ${\psi}^{(j,\beta)}_{n}(x)$, like $\psi^{\bt}_{n}(x)$
satisfy Eqs.(\ref{recursion4}) and (\ref{recursion5})
respectively.

Similarly, for $\beta=4$, we get for $n\geq 2d$,
\begin{eqnarray}
\label{beta4tilde} \nonumber x\frac{d}{dx}{\phi}^{(j,4)}_n(x)
&=& x{\psi}^{(j,4)}_n(x), \\
&=& \sum R^{(4)}_{n,m}{\phi}^{(j,4)}_m(x),
\end{eqnarray}
where the first term drops off due to the condition $n\geq 2d$,
$d\geq 1$. For $\beta=1$, we differentiate (\ref{ddxtildepsi}) to
get
\begin{eqnarray}
x\frac{d}{dx}{\phi}^{(j,1)}_n(x) &=& \sum
(R^{(1)}_{nm}-\delta_{n,m}){\phi}^{(j,1)}_n(x),\qquad n\geq 2d.
\end{eqnarray}
This in turn proves that for $n\geq 2d$, these functions satisfy
the same set of recursion relations (\ref{difference1}) and
(\ref{difference2}) as done by the skew-orthogonal vectors.

Thus we prove that our function is compatible under a system of
difference-differential equation, i.e. the solution of the system
of differential equation is also a simultaneous solution of the
system of difference equation. We will now show that these
solutions are also compatible with the system of deformation
relations i.e. they are also solutions of the system of PDE's.

We can see that
\begin{eqnarray}
\label{psitildeu} \nonumber \frac{\partial}{\partial
u_K}{\psi}^{(j,\beta)}_n(x) &=& \frac{f_j(x)}{K}\int_{\mathbb
R}\frac{x^K-z^K}{x-z}\psi^{\bt}_{n}(z)\exp[-V(z)]dz
+\sum {(U_K^{\beta})}_{nm}{\psi}^{(j,\beta)}_m(x),\\
&=& f_j(x)\sum_{m=0}^{k-1}a_{m}^{\bt}(x)Z_{m,n}+\sum
{(U_K^{\beta})}_{n,m}{\psi}^{(j,\beta)}_m(x).
\end{eqnarray}
Thus for $n\geq 2d$, ${\psi}^{(j,\beta)}_n(x)$ satisfy
Eq.(\ref{d2}). Also, combining
Eqs.(\ref{phitildeprime},\ref{psitildeu}), we can see that

\begin{eqnarray}
{{\phi}}^{(j,\beta)}_n(x) &=&\sum
{(U_K^{\beta})}_{n,m}{\phi}^{(j,\beta)}_m(x),\qquad n\geq 2d.
\end{eqnarray}

{\it Remark:} Here we encounter  terms of the form $f_{j+K}(x)$,
$j=0,\ldots, 2d-2$ and $K=1,\ldots, 2d$. For $(j+K)>2d-2$, this
term does not contribute, by definition. Also if we consider only
$K=2d$, i.e. the leading term in $V(x)$, no such problem arise.

Thus for $n \geq 2d$, the new-functions ${\phi}_n^{(j,\beta)}(x)$
and ${\psi}_n^{(j,\beta)}(x)$ satisfy the system of
differential-difference-deformation equation and hence are
mutually compatible.

Explicitly, we have
\begin{eqnarray}
\nonumber {\Psi_{[W]}^{\bt}} =
 \left(\begin{array}{cccccc}
{\Psi}_{N-d}^{(\beta)}(x) & {\tilde\Psi}_{N-d}^{(\beta)}(x) &
{\Psi}_{N-d}^{(0,\beta)}(x)
            & \ldots & {\Psi}_{N-d}^{(2d-2,\beta)}(x)    \\
\vdots          & \vdots                              & \vdots                 & \vdots & \vdots    \\
{\Psi}_{N+d-1}^{(\beta)}(x) & {\tilde\Psi}_{N+d-1}^{(\beta)}(x) &
{\Psi}_{N+d-1}^{(0,\beta)}(x)
            & \ldots & {\Psi}_{N+d-1}^{(2d-2,\beta)}(x)    \\
\end{array}\right),\\
{\Phi_{[W]}^{\bt}} =
 \left(\begin{array}{ccccccc}
{\Phi}_{N-d}^{(\beta)}(x) & {\tilde\Phi}_{N-d}^{(\beta)}(x) &
{\Phi}_{N-d}^{(0,\beta)}(x)
            & \ldots & {\Phi}_{N-d}^{(2d-2,\beta)}(x)    \\
\vdots          & \vdots                       & \vdots                 & \vdots & \vdots    \\
{\Phi}_{N+d-1}^{(\beta)}(x) & {\tilde\Phi}_{N+d-1}^{(\beta)}(x) &
{\Phi}_{N+d-1}^{(0,\beta)}(x) & \ldots & {\Phi}_{N+d-1}^{(2d-2,\beta)}(x)    \\
\end{array}\right).
\end{eqnarray}

Thus we have obtained the fundamental solution to the system of
differential-difference-deformation equation, satisfied by the
skew-orthogonal vectors. From here, it should be possible to
define the RHP for these skew-orthogonal vectors. It has been
obtained in Ref.\cite{pierce} for $\beta=1$ and can be extended to
$\beta=4$.

\section{Compatibility of the Finite Difference-Differential-Deformation Systems}

In the previous section, we have obtained the Fundamental system
of solutions for the differential-difference-deformation equation.

In this section, we will provide an alternate proof to show that
the recursion relations, the linear differential equations and the
deformation equations are compatible in the sense that they admit
a basis of simultaneous solutions, provided the vectors form a
certain algebra defined by the commutation relations
(\ref{string}) and (\ref{string2}). In other words, here, we will
not use the explicit form of the fundamental solution to prove the
compatibility.

This is the same as saying that the shifts $W \mapsto W+1$ in
Eqs.(\ref{recursion1}), (\ref{recursion2}) respectively, are
compatible as vector differential-difference systems. This means
that there exists a sequence of fundamental matrix solutions
${\bf\Phi^{\bt}_{[W]}}(x)$ and ${\bf\Psi^{\bt}_{[W]}}(x)$ which
are simultaneous solutions of
\begin{eqnarray}
{\bf\Phi^{(4)}_{[W+1]}}(x)&=& A_N^{(4)}(x){\bf\Phi^{(4)}_{[W]}}(x),\\
\frac{d}{dx}{\bf\Phi^{(4)}_{[W]}}(x)&=&
D_N^{(4)}(x){\bf\Phi^{(4)}_{[W]}}(x),
\end{eqnarray}
and
\begin{eqnarray}
{\bf\Psi^{(4)}_{[W+1]}}(x)&=& B_N^{(4)}(x){\bf\Psi^{(4)}_{[W]}}(x),\\
\frac{d}{dx}{\bf\Psi^{(4)}_{[W]}}(x)&=& {\underline
D}_N^{(4)}(x){\bf\Psi^{(4)}_{[W]}}(x),
\end{eqnarray}
respectively. The same result holds for $\beta=1$. This means that
there exists a sequence of fundamental matrix solutions satisfying
the recursion and differential equations

\begin{eqnarray}
{\bf\Psi^{(1)}_{[W+1]}}(x)&=&
A_N^{(1)}(x){\bf\Psi^{(1)}_{[W]}}(x),\\
\frac{d}{dx}{\bf\Psi^{(1)}_{[W]}}(x)&=&
D_N^{(1)}(x){\bf\Psi^{(1)}_{[W]}}(x),
\end{eqnarray}
and
\begin{eqnarray}
{\bf\Phi^{(1)}_{[W+1]}}(x)&=&
B_{N}^{(1)}(x){\bf\Phi^{(1)}_{[W]}}(x),\\
\frac{d}{dx}{\bf\Phi^{(1)}_{[W]}}(x)&=&  {\underline
D}_N^{(1)}(x){\bf\Phi^{(1)}_{[W]}}(x),
\end{eqnarray}
respectively.

One must note that the differential equations are also compatible
with the shift $W\mapsto W-1$. The proof is exactly similar to the
above case.

Similar procedure will be repeated to show the compatibility
between the difference-deformation equation.


\subsection{Proof: Difference-differential equation.}
Now we will prove the compatibility of the differential-difference
equation for  $\Phi^{(4)}(x)$. The others are exactly similar and
hence not repeated. The proof follows on similar line to the one
outlined for bi-orthogonal polynomials \cite{eynard2}.

Let us define a finite subsequence (or window) containing
functions:
\begin{equation}
{\overline\Phi}_{W}^{(4)}(x):={\left[{\overline\Phi}_{N-d}^{{(4)}^t}(x),
\ldots, {\overline\Phi}_{N+d-1}^{{(4)}^t}(x)\right]}^{t},\qquad
N\geq d,
\end{equation}
which is a solution to the differential equation
\begin{eqnarray}
\frac{d}{dx}{{\overline\Phi}^{(4)}_{W}}(x)=
D_N^{(4)}(x){{\overline\Phi}^{(4)}_{W}}(x).
\end{eqnarray}
A glance through the window will confirm that the shift
$W\longmapsto W+1$ will introduce one new function
${\overline\Phi}_{N+d}^{(4)}(x)$ inside the shifted window.
Componentwise, let us suppose that they satisfy the recursion
relation
\begin{equation}
\label{recursion3}
\left[\eta_{-d}(m)-x\zeta_{-d}(m)\right]{\overline\Phi}_{m+d}^{(4)}(x)=
x\sum_{l=-d+1}^{d}\zeta_{l}(m){\overline\Phi}_{m-l}^{(4)}(x)
-\sum_{l=-d+1}^{d}\eta_{l}(m){\overline\Phi}_{m-l}^{(4)}(x),\qquad
m\geq d.
\end{equation}
It is easy to see that the differential equation componentwise
reads:
\begin{equation}
\frac{d}{dx}{\overline\Phi}_{n}^{(4)}(x)
=\sum_{l=-d}^{d}\zeta_{j}(n) {\overline\Phi}_{n-l}^{(4)}(x),\qquad
n=N-d,\ldots N+d-1,
\end{equation}
where the ${\overline\Phi}_{m}^{(4)}(x)$ outside the window can be
expressed in terms of that within the window by using
Eq.(\ref{recursion3}) recursively. To show compatibility between
the shift and differential operator, we need to show that the
newly defined function, which is assumed to satisfy
\begin{equation}
{\overline\Phi}_{N+d}^{(4)}(x)={(\eta_{-d}(N))}^{-1}\left[
x\sum_{l=-d}^{d}\zeta_{l}(N){\overline\Phi}_{N-l}^{(4)}(x)
-\sum_{l=-d+1}^{d}\eta_{l}(N){\overline\Phi}_{N-l}^{(4)}(x)\right],
\end{equation}
will also satisfy  the same differential equation  i.e. we must
show that the newly defined function
${\overline\Phi}_{N+d}^{(4)}(x)$ will also be a solution to the
differential equation:
\begin{equation}
\label{psii}
\frac{d}{dx}{\overline\Phi}_{N+d}^{(4)}(x)=\sum_{l=-d}^{d}\zeta_{j}(N+d){\overline\Phi}_{N-l+d}^{(4)}(x).
\end{equation}
{\it Remark.} One must be careful about the position of the
inverse of the quaternion, as in general, they do not commute.

Once this is proved, we can argue by induction that
${\overline\Phi}_{N+d+j}^{(4)}(x)$ satisfy the same sort of
differential equation for any $j>1$. To do this, we compute
\begin{eqnarray}
\nonumber
\eta_{-d}(N)\left(\frac{d}{dx}\right){\overline\Phi}_{N+d}^{(4)}(x)
&=&
\frac{d}{dx}\left(x\sum_{l=-d}^{d}\zeta_{l}(N){\overline\Phi}_{N-l}^{(4)}(x)
-\sum_{l=-d+1}^{d}\eta_{l}(N){\overline\Phi}_{N-l}^{(4)}(x)\right)\\
&=& \nonumber\label{final}
\sum_{l=-d}^{d}\zeta_{l}(N){\overline\Phi}_{N-l}^{(4)}(x) 
+\sum_{l=-d}^{d}\sum_{j=-d}^{d}{\zeta}_{l}
(N){\eta}_{j}(N-l){\overline\Phi}_{N-l-j}^{(4)}(x)\\
\nonumber &&
-\sum_{l=-d}^{d}\sum_{j=-d}^{d}{\eta}_{l}(N)\zeta_{j}(N-l){\overline\Phi}_{N-l-j}^{(4)}(x)\\
&& + \eta_{-d}(N)\sum_{l=-d}^{d}\zeta_{l}(N+d){\overline\Phi}_{N-l+d}^{(4)}(x). 
\end{eqnarray}
From Eq.(\ref{final}), we can see that in order to prove Eq.
(\ref{psii}), we will have to show that
\begin{eqnarray}
\sum_{l=-d}^{d}\zeta_{l}{\overline\Phi}_{N-l}^{(4)}(x) 
+\sum_{l=-d}^{d}\sum_{j=-d}^{d}{\zeta}_{l}
(N){\eta}_{j}(N-l){\overline\Phi}_{N-l-j}^{(4)}(x)
-\sum_{l=-d}^{d}\sum_{j=-d}^{d}{\eta}_{l}(N)\zeta_{j}(N-l){\overline\Phi}_{N-l-j}^{(4)}(x)=0.
\end{eqnarray}
But this is nothing but a direct consequence of
Eq.(\ref{string1}), in terms of its components. We can repeat the
same procedure for the shift $W\longmapsto W-1$. Finally, one can
extend this argument by induction to
${\overline\Phi}_{N+r}^{(4)}(x)$, thereby completing the proof.

We can repeat the same procedure to the sequences
\begin{equation}
{\overline\Psi}_{W}^{(1)}(x):={\left[{\overline\Psi}_{N-d}^{{(1)}^t}(x),
\ldots, {\overline\Psi}_{N+d-1}^{{(1)}^t}(x)\right]}^{t},\qquad
N\geq d,
\end{equation}
and with some minor modifications to
\begin{equation}
{\overline\Phi}_{W}^{(1)}(x):={\left[{\overline\Phi}_{N-d}^{{(1)}^t}(x),
\ldots, {\overline\Phi}_{N+d-1}^{{(1)}^t}(x)\right]}^{t},\qquad
N\geq d,
\end{equation}
and \begin{equation}
{\overline\Psi}_{W}^{(4)}(x):={\left[{\overline\Psi}_{N-d}^{{(4)}^t}(x),
\ldots, {\overline\Psi}_{N+d-1}^{{(4)}^t}(x)\right]}^{t},\qquad
N\geq d,
\end{equation}
to prove compatibility conditions between the
difference-differential equations satisfied by the
skew-orthonormal vectors.

\subsection{Proof: Difference-deformation equation.}

We show that these functions are also compatible with the
deformation equations. We will briefly outline the proof. We start
with the same  finite subsequence (or window) containing
functions:
\begin{equation}
{\overline\Phi}_{W}^{(4)}(x):={\left[{\overline\Phi}_{N-d}^{(4)}(x),
\ldots, {\overline\Phi}_{N+d-1}^{(4)}(x)\right]}^{t},\qquad N\geq
d,
\end{equation}
which componentwise, will satisfy the PDE:
\begin{equation}
\frac{\partial}{\partial
u_{K}}{\overline\Phi}_{n}^{(4)}(x)=\sum_{j=0}^{K}U^{K}_{j}(n){\overline\Phi}_{n-j}^{(4)}(x)\qquad
n=N-d,\ldots,N+d-1.
\end{equation}
Then under a shift $W\longmapsto W+1$, we get the new function
${\overline\Phi}_{N+d}^{(4)}(x)$ defined by
\begin{equation}
(\eta_{-d}(N)){\overline\Phi}_{N+d}^{(4)}(x)=x\left[
\sum_{l=-d}^{d}\zeta_{l}(N){\overline\Phi}_{N-d+1}(x)-\sum_{l=-d+1}^{d}\eta_{l}(N){\overline\Phi}_{N-l}^{(4)}(x)\right],
\end{equation}
which should satisfy
\begin{equation}
\label{psiii} \frac{\partial}{\partial
u_{K}}{\overline\Phi}_{N+d}^{(4)}(x)=\sum_{j=0}^{K}U^{K}_{j}(N+d){\overline\Phi}_{
N+d-j}^{(4)}(x).
\end{equation}
To do this, we compute
\begin{eqnarray}
\nonumber \left(\frac{\partial}{\partial
u_{K}}\right)(\eta_{-d}(N)){\overline\Phi}_{N+d}^{(4)}(x) &=&
\frac{\partial}{\partial u_{K}}\left(x\left[
\sum_{l=-d}^{d}\zeta_{l}(N){\overline\Phi}_{N-l}^{(4)}(x)
-\sum_{l=-d+1}^{d}\eta_{l}(N){\overline\Phi}_{N-l}^{(4)}(x)\right]\right),\\
\nonumber &=&x\sum_{l=-d}^{d}{\zeta}'_{l}
(N){\overline\Phi}_{N-l}^{(4)}(x)+ 
x\sum_{l=-d}^{d}\sum_{j=0}^{K}{\zeta}_{l}
(N)U^{K}_{j}(N-l){\overline\Phi}_{N-j-l}^{(4)}(x)\\
&& \nonumber
 -\sum_{j=-d}^{d}{\eta}'_{l}(N){\overline\Phi}_{N-l}^{(4)}(x)
-\sum_{l=-d}^{d}\sum_{j=0}^{K}\eta_{l}(N)U^{K}_{j}(N-l){\overline\Phi}_{N-j-l}^{(4)}(x)\\
&&
+\eta_{-d}(N)\sum_{j=0}^{K}U^{K}_{j}(N+d){\overline\Phi}_{N+d-j}^{(4)}(x)
+{\eta}'_{-d}(N){\overline\Phi}_{N+d}^{(4)}(x),
\end{eqnarray}
where ${}^{\prime}$ denotes $\frac{\partial}{\partial u_K}$.
Rearranging the coefficients in Eq(\ref{psii}), we can see that
basically we will have to show that
\begin{eqnarray}
\nonumber 0&=& x\sum_{l=-d}^{d}{\zeta}'_{l}
(N){\overline\Phi}_{N-l}^{(4)}(x)+ 
x\sum_{l=-d}^{d}\sum_{j=0}^{K}{\zeta}_{l}
(N)U^{K}_{j}(N-l){\overline\Phi}_{N-j-l}^{(4)}(x)\\
&&
 -\sum_{j=-d}^{d}{\eta}'_{l}(N){\overline\Phi}_{N-l}^{(4)}(x)
-\sum_{l=-d}^{d}\sum_{j=0}^{K}\eta_{l}(N)U^{K}_{j}(N-l){\overline\Phi}_{N-j-l}^{(4)}(x).
\end{eqnarray}
But this is nothing but a direct consequence of the string
Eq.(\ref{string2}), in terms of its components. Hence we get
Eq.(\ref{psiii}).

We can repeat the same procedure for the shift $W\longmapsto W-1$.
Finally, one can extend this argument by induction to
${\overline\Phi}_{N+r}^{(4)}(x)$, thereby completing the proof.

We can repeat the same procedure to the sequences
${\overline\Psi}_{W}^{(1)}(x)$ and with some minor modifications
to ${\overline\Phi}_{W}^{(1)}(x)$ and
${\overline\Psi}_{W}^{(4)}(x)$ to prove compatibility conditions
between the difference-deformation equations satisfied by the
skew-orthonormal vectors.

\section{Conclusion}

In conclusion, we have obtained a system of
differential-difference-deformation equation for a finite
subsequence of skew-orthonormal vectors. We see that similar to
the orthogonal and bi-orthogonal polynomials in ordinary space,
the Cauchy-like transforms of these skew-orthogonal vectors, of
sufficiently high order (to be precise, $N\geq 2d$) also satisfy
the same differential-difference-deformation equations in the
quaternion space. We also derive an integral representation of
these skew-orthogonal vectors in order to obtain the fundamental
system of solutions for the overdetermined system of ODEs,
difference and deformation equations.

On the other hand, from our already existing knowledge of
skew-orthogonal polynomials, we know that the final results of the
RH-analysis will have some distinct differences from the
orthogonal polynomials. For example, the zeros of orthogonal
polynomials are real, which is in contrast to that of the
skew-orthogonal polynomials \cite{ghoshpandey}. To understand
these properties, one needs to extend the already existing theory
for orthogonal polynomials and matrix-RHP \cite{deift4} to
skew-orthogonal polynomials and q-matrix RHP. We wish to come back
to this in a later publication.

Finally, we would like to mention a result of Dyson and Mehta.
They showed \cite{dysommehta} that for the circular ensembles,
``the probability distribution of a set of $N$ alternate
eigenvalues of a matrix in the Orthogonal ensemble of order $2N$
is identical with the probability distribution of the set of all
eigenvalues of a matrix in Symplectic ensemble of order $N$''. It
would be interesting to know if something similar exist for
ensembles with  polynomial potential and the role (if any) played
by the duality relations between different skew-orthogonal vectors
of Orthogonal and Symplectic ensembles.

\section{Acknowledgment} I am grateful to Bertrand Eynard for
teaching me techniques related to the two-matrix model which have
been extremely useful in this paper. I also acknowledge the
referee for some extremely important hints and suggestions,
specially in section 8 of the article.

\end{document}